\documentclass[CRPHYS,Unicode,manuscript]{cedram}

\hypersetup{
    colorlinks=true,
    linkcolor=blue,
    filecolor=magenta,      
    urlcolor=cyan,
    pdftitle={Overleaf Example},
    pdfpagemode=FullScreen,
    }
\definecolor{blue}{rgb}{0,0,1}
\title{%
Study of atmospheres in the solar system, from stellar occultation or planetary transit \\
Etude des atmosph\`eres au sein du syst\`eme solaire, par occultations stellaires ou transits plan\'etaires
}%

\author{\firstname{Bruno} \lastname{Sicardy}\CDRorcid{0000-0003-1995-0842}\IsCorresp}  
\address{%
LESIA, Observatoire de Paris,
Universit\'e PSL, CNRS, Sorbonne Universit\'e,
Universit\'e de Paris, 5 place Jules Janssen, 92195 Meudon, France
}%
\email[B. Sicardy]{bruno.sicardy@obspm.fr}
\thanks{
The work leading to these results has received funding from the 
European Research Council under the European Community's H2020
2014-2021 ERC Grant Agreement no. 669416 ``Lucky Star".
}% 
\ESM{%
DOI
}%
\keywords{%
Refraction, Planetary Atmospheres, Stellar Occultations, Transits,
R\'efraction, Atmosph\`eres Plan\'etaires, Occultations Stellaires, Transits,
}%

\begin{abstract}
Stellar occultations and transits occur when a planetary body passes in front of a star (including our Sun).
For objects with an atmosphere, refraction plays an essential role to explain 
the drops of flux and the aureoles observed during these events.
This can be used to derived key parameters of the atmospheres, such as their
density, pressure and temperature profiles, as well as the presence of atmospheric gravity waves and zonal winds.
Here we derive from basic principles the equations that rule the ray propagation in planetary atmospheres,
and we show how they can be used to derive the physical parameters of these atmospheres. \\
Les occultations stellaires et les transits se produisent lorsqu'un corps plan\'etaire passe devant une \'etoile (y compris notre Soleil).
Pour les objets avec une atmosph\`ere, le r\^ole  de la r\'efraction est essentiel pour expliquer 
les chutes de flux et les aur\'eoles observ\'ees lors de ces \'ev\'enements.
Ces derniers peuvent \^etre utilis\'es pour d\'eduire des param\`etres cl\'es des atmosph\`eres, comme leurs
profils de densit\'e, de pression et de temp\'erature, ainsi que la pr\'esence d'ondes de gravit\'e ou de vents zonaux.
A partir des principes fondamentaux, nous d\'eduisons les \'equations qui r\'egissent la propagation des rayons 
dans les atmosph\`eres plan\'etaires, et nous montrons comment elles peuvent \^etre utilis\'ees pour d\'eduire 
les param\`etres physiques de ces atmosph\`eres.
\end{abstract}

\begin{document}

% Use the \maketitle command after the abstract
\maketitle

\section{Introduction}

As planetary bodies move in space, they may pass in front of another object, 
as seen from an observer on Earth or from an instrument on board a spacecraft.
Different terminologies are used to describe these phenomena. 
An \textit{occultation} occurs when a body blocks the light from a background object.
A typical example is a stellar occultation, where a planetary body passes in front of a star.
In these cases, the physical disk of the star usually appears as much smaller than the size of the occulting body itself.
For instance, the angular diameter of a star projected at the typical distances of an asteroid 
amounts to a kilometer at most, while the asteroid itself may have a diameter of tens of kilometers.

Conversely, \textit{transits} occur when the foreground object is angularly much smaller than the background object. 
Famous examples are transits of Mercury or Venus in front of the Sun as seen from Earth.
In the last two decades, transits of exoplanets in front of their stars have been a very powerful tool to 
discover exoplanets, assess their sizes and orbital periods, 
measure perturbations from other exoplanets around the same star, 
or detect chemical species in their atmospheres.

Finally, the term \textit{eclipse} refers to a body casting its shadow on another body.
A well known  example is given by lunar eclipses\footnote{In this context,
solar eclipses should actually be referred to as ``solar occulations".}.
While eclipses of the Galilean satellites have been observed since their discovery by Galileo,
these phenomena among satellites of Jupiter and Saturn have been widely observed In the last decades
to pin down their orbital elements and assess secular trends caused for instance by tidal effects.

Here we focus on stellar occultations and transits involving bodies with atmospheres.
During a stellar occultation by an opaque object, 
the star abruptly disappears when reaching the limb of that object.
More precisely, the sharpness of the disappearance and reappearance of the star 
are only limited by the stellar diameter and diffraction effects, 
and typically last for a fraction of a second only.

In contrast, occultations by objects with an atmosphere are gradual and may last for several minutes.
In fact, even if the occultation is diametric, 
the occulted star may remain faintly visible during the entire event, 
due to the refraction of the stellar rays by the atmosphere.
Contrarily to what is often thought, the gradual character of atmospheric occultations 
is usually \textit{not} caused by absorption (due for instance to hazes).
In fact, even a completely transparent atmosphere can cause the gradual disappearance 
of the star (or on the contrary, its brightening), through refraction effects.

In fact, the atmosphere acts as a lens that may focus or defocus the stellar flux.
As we will see, the observed phenomena bear some ressemblance with gravitational
lenses, where the ray bending stems from gravity instead of gas refraction. 

\section{The equations of refraction from basic principles}

We recall here a few basic principles that provide the equations of propagation of a luminous ray in an atmosphere.
The first principle was stated by Fermat, who noted that during its propagation between two points $A$ and $B$,
a luminous ray  minimizes the time of travel between $A$ and $B$. Since the velocity of light is 
$v= c/n$, where $c$ is the speed of light in the vacuum and $n$ is the index of refraction, Fermat's principle
may be expressed as the fact that the optical path $l$, where 
\begin{equation}
l = \int_A^B n ds,
\label{eq_l}
\end{equation}
is stationary (and usually minimal) during the ray propagation.

Fermat's principle can be derived from the undulatory nature of light (Huygens' principle), and
more generally, from the more modern principle of least action, widely used in Quantum 
Mechanics\footnote{See the book  by Feynman, where the connection between light and quantum electrodynamics
(QED) is described for a informed but wide public \cite{fey85}.}.
In practice, the problem of minimizing the integral $l$ above is solved by the classical Euler and Lagrange equation,
which provides (\cite{goo95} p. 1.20-24)
\begin{equation}
\frac{d \vec{\tau}}{ds} =  \vec{\nabla} n,
\label{eq_dtau_ds}
\end{equation}
with $\vec{\tau}= n \hat{u}$, 
where $\hat{u}$ is the unit vector tangent to the ray at $\vec{r}$ and $\vec{\nabla} n$ is the gradient of $n(\vec{r})$.

In the case where the ray propagates in a plane, the equation above can be written in another way.
During an elementary displacement $ds$, the ray is deflected by an elementary angle $d\omega$,
so that $\hat{u}$ suffers a deviation $d\omega$ given by
\begin{equation}
d \hat{u}= \hat{v} d\omega,
\label{eq_du_domega}
\end{equation}
where $\hat{v}$ is the unit vector perpendicular to $\hat{u}$.
Using Eqs.~\ref{eq_dtau_ds} and \ref{eq_du_domega}, we obtain after elementary calculations
\begin{equation}
\frac{d\omega}{ds} 
= \left( \frac{\vec{\nabla} n}{n}  \right) \cdot \hat{v},
%= \vec{\nabla} [ \ln(n) ] \cdot \hat{v}.
\label{eq_domega_ds}
\end{equation}
which will be used in the rest of this chapter.
Note that Eq.~\ref{eq_domega_ds} is equivalent to Eq.~\ref{eq_dtau_ds} only in the case of \it a planar propagation of the ray. \rm
This is not true anymore for a 3D propagation, where the torsion of the ray must also be accounted for.

\section{Refraction by planetary atmospheres}
\label{sec_refrac_plan_atmo}

The effects of refraction by a planetary atmosphere during Venus transits have been studied quite long ago,
with articles dating back to the eighteenth and nineteenth centuries, see Section~\ref{sec_transits}.
Applications to stellar occultations, on the other hand, started to be discussed one century ago or so,
in particular by Anton Pannekoek \cite{pan03} in 1903 and by Charles Fabry \cite{fab29} in 1929 .

Observations of stellar occultations were difficult at that time, though, 
because they required a fast and sensitive photometric recording device.
Only on 20 November 1952 was an occultation of the star $\sigma$ Arietis by Jupiter recorded \cite{bau53}. 
Another occultation was monitored on 7 July 1959, involving Venus that passed that time in front of the 
bright star $\alpha$ Leonis (alias Regulus) \cite{dev60}.
More than one decade was necessary to have another event recorded, on 13 May 1971, 
when Jupiter occulted the star $\beta$ Scorpii \cite{vev72,vap73,was73}.

\begin{figure}[!h]
\centering
\includegraphics[totalheight=4cm,trim=0 0 0 0]{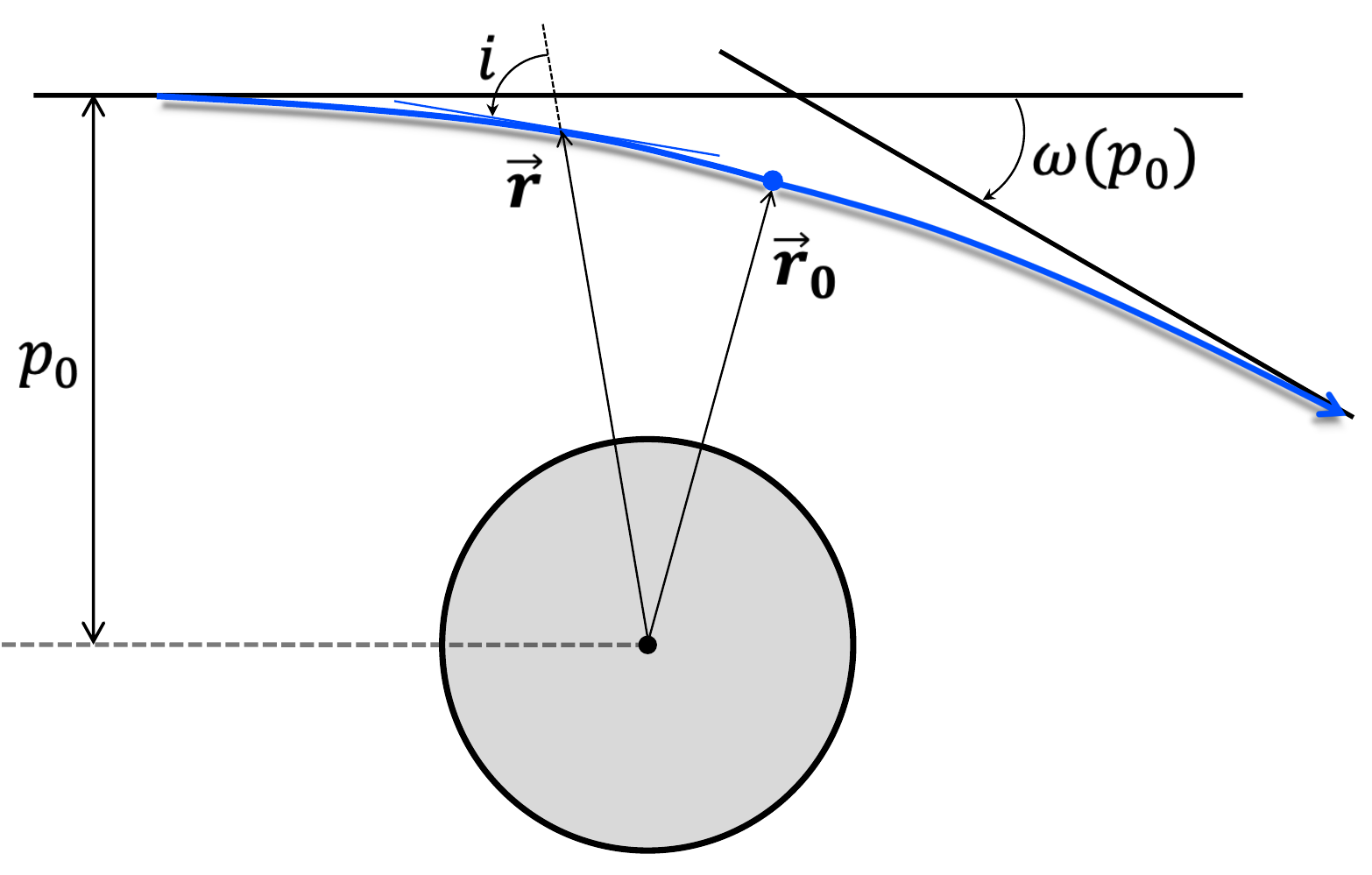}
\caption{%
A ray coming from infinity from the left with impact parameter $p_0$ is refracted by the atmosphere of a planet.
It reaches its closest approach at  $\vec{r}_0$ and goes to infinity again with a total
deviation of $\omega(p_0)$ that depends only on the incoming impact parameter.
Note that by convention here, the angle $\omega$ is negative while $i$ is positive.
}%
\label{fig_deviation_total}
\end{figure}

The formal equations describing the effects of atmospheres during occultations are found
in various works from the beginning of the 1970's \cite{fje71,vap73,was73}.
They consider the simplest possible case: a planet with a spherically symmetric and transparent atmosphere. 
Its refractive index $n(r)$ then depends only on the distance $r$ to the planet's center.
A ray coming from infinity from the left with impact parameter $p_0$
is refracted by the atmosphere and returns to infinity at the right with a total deviation $\omega(p_0)$, 
after passing at a closest distance $r_0(p_0)$ from the planet's center (Fig.~\ref{fig_deviation_total}).
From the symmetry of the problem, the ray propagates in the plane that passes through the center of
symmetry of the atmosphere, as depicted in Fig.~\ref{fig_deviation_total}.

%%%%%%%%%%%%%%%%%%%%%%%%%%%%%%%%%%%%%%%%%%
%%%%%%%%%%%%%%%%%%%%%%%%%%%%%%%%%%%%%%%%%%

\subsection{Bouguer's rule}

Let us denote $i$ the angle between the position vector $\vec{r}$ of the current point along the ray path
and the tangent to the path, see Fig.~\ref{fig_deviation_total}. We have
$$
\sin(i) = \left( \hat{u} \times \hat{r} \right) \cdot \hat{z},
$$
where 
$\hat{u}$ is the unit vector in the direction of the ray propagation,
$\hat{r}$ is the unit radial vector $\hat{r} = \vec{r}/r$, and
$\hat{z}$ is the unit vector perpendicular to the plane of Fig.~\ref{fig_deviation_total}.
Using $\vec{\tau}= n \hat{u}$, we define $B$ as
$$
B= n r \sin(i) =  \left( \vec{\tau} \times \vec{r} \right) \cdot \hat{z}.
$$
As the ray progresses by a displacement $ds$ along the path, $B$ varies at the rate
$$
\frac{dB}{ds} = 
\left( \frac{d\vec{\tau}}{ds} \times \vec{r} \right) \cdot \hat{z} +
\left( \vec{\tau} \times \frac{d\vec{r}}{ds} \right) \cdot \hat{z}= 
\left( \vec{\nabla} n \times \vec{r}  \right) \cdot \hat{z} +
\left( \vec{\tau} \times \hat{u} \right) \cdot \hat{z}= 0.
$$
This directly results from the fact that 
the atmosphere is spherically symmetric, so that $\vec{\nabla} n$ is parallel to $\vec{r}$,
and from the fact that $\vec{\tau}$ is parallel to $\hat{u}$ by definition. 
We then obtain the \it Bouguer's rule, \rm $dB/ds=0$, i.e.
$$
n r \sin(i)= {\rm~constant~along~the~path.}
$$
Very far away from the planet (at left in Fig.~\ref{fig_deviation_total}), there is no atmosphere ($n=1$) and $r \sin(i) = p_0$.
At closest approach, $i= \pi/2$ and $B= n_0 r_0$, where $n_0= n(r_0)$.
Hence
\begin{equation}
n r \sin(i) = n_0 r_0 = p_0,
\label{eq_bouguer_rule}
\end{equation}
which relates the impact parameter $p_0$, 
the closest approach distance $r_0$ of the ray and
the index of refraction $n_0$ at that point.
%
% Note that Bouguer's rule is the equivalent for a spherical atmosphere of 
% the Snell-Descartes' law $n \sin(i)=$ constant that applies for a horizontally layered medium. 

\subsection{Total deviation of the ray}

Since the ray propagates in a plane in the case examined here, we can use Eq.~\ref{eq_domega_ds}
to calculate the ray deviation $\omega$ as the ray progresses in the atmosphere.
For a spherically symmetric atmosphere, $\vec{\nabla} n$ can be expressed as
$$
\vec{\nabla} n = \frac{dn}{dr} \hat{r} = \frac{1}{r} \frac{dn}{dr} \vec{r}.
$$
Hence
$$
\frac{d\omega}{ds} = \frac{1}{n} \frac{dn}{dr} \left( \hat{r} \cdot \hat{v} \right).
$$ 
%
%%%%%%%%%%%%%%%%%%%%%%%%%%%%%%%%%%%%%%%%%%
\begin{figure}[!b]
\centering
\includegraphics[totalheight=4cm,trim=0 0 0 0]{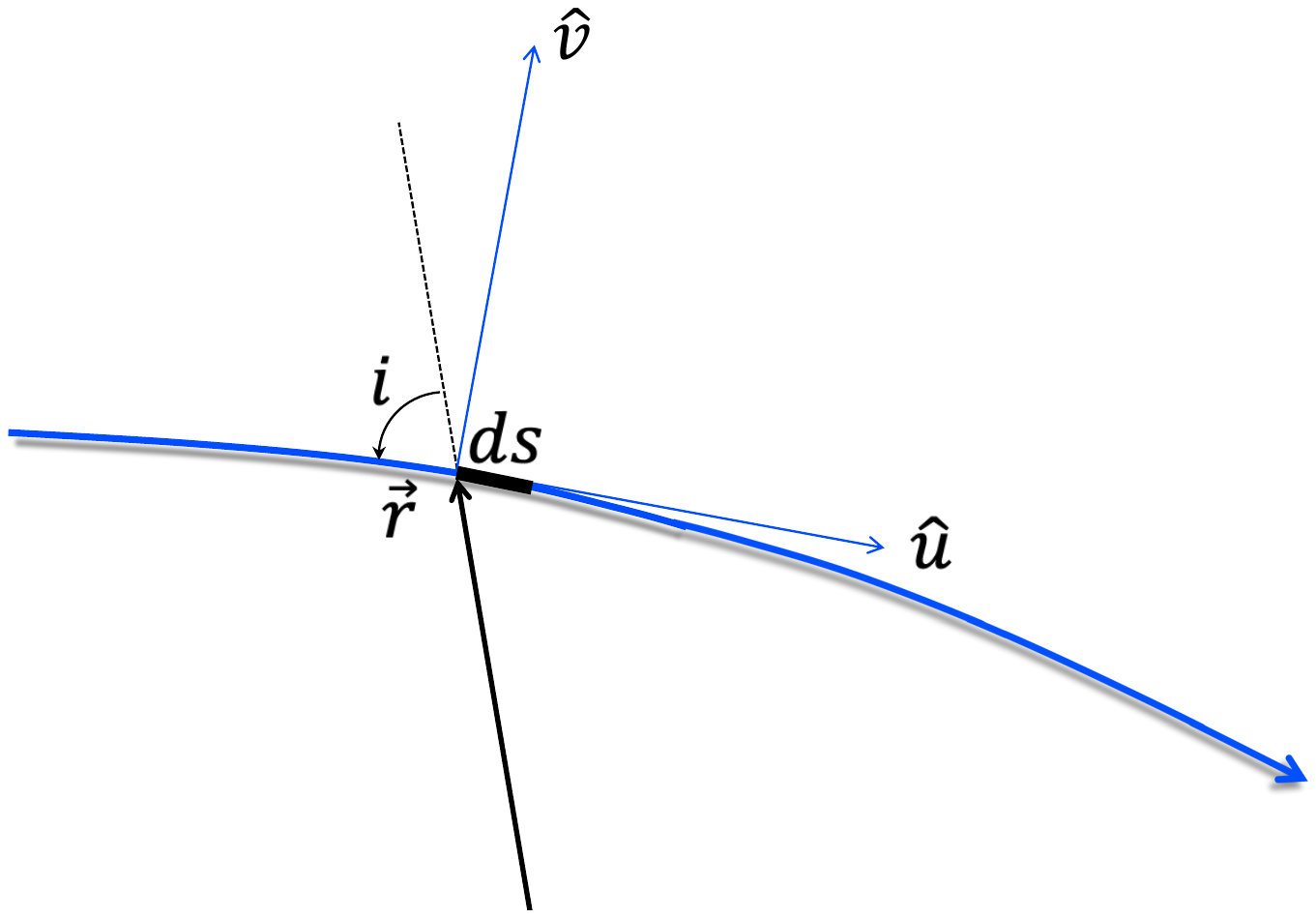}
\caption{%
The definitions of the vectors and angles used in the text.
The ray is in blue.
}%
\label{fig_deviation_local}
\end{figure}
%%%%%%%%%%%%%%%%%%%%%%%%%%%%%%%%%%%%%%%%%%
%
We see in Fig.~\ref{fig_deviation_local} that $\hat{r} \cdot \hat{v}= \sin(i)$,
where $\sin (i) = p_0/rn$ (Eq.~\ref{eq_bouguer_rule}), thus
$$
\frac{d\omega}{ds} = \frac{p_0}{r n^2} \frac{dn}{dr}.
$$
Moreover, again from Fig.~\ref{fig_deviation_local} we see that $dr= -\cos (i) ds$.
From $\cos (i)= \sqrt{1-\sin^2 (i)}$ and $\sin (i) = p_0/rn$, we finally arrive at
$$
\frac{d\omega}{dr} = -\frac{p_0}{n} \frac{dn}{dr} \frac{1}{\sqrt{r^2 n^2 - p_0^2}}.
$$
By integrating $d\omega/dx$ from $+\infty$ (corresponding to a ray coming from infinity at the left of Fig.~\ref{fig_deviation_total}) 
to $r_0$ (the closest approach to the planet), we obtain half of the total deviation, 
so that the total deviation is 
\begin{equation}
\omega (p_0)= 2 \int_{r_0(p_0)}^{+\infty} \frac{p_0}{n} \frac{dn}{dr}  \frac{dr}{\sqrt{r^2 n^2 - p_0^2}},
\label{eq_total_devia}
\end{equation}
where we recall that $p_0= r_0 n_0$.

Suppose that we know the refractive structure of the atmosphere, i.e. the profile $n(r)$.
Then, for each impact parameter $p_0$, we can determine the closest approach distance
$r_0$ by solving the equation $r_0 n(r_0)= p_0$. 
This entirely determines $\omega (p_0)$ using a numerical scheme to calculate the integral above.
In other words, to each impact parameter $p_0$, we can associate a deviation angle $\omega (p_0)$.

\section{Retrieval of the atmospheric structure}

\begin{figure}[!h]
\centering
\includegraphics[totalheight=4cm,trim=0 50 0 50]{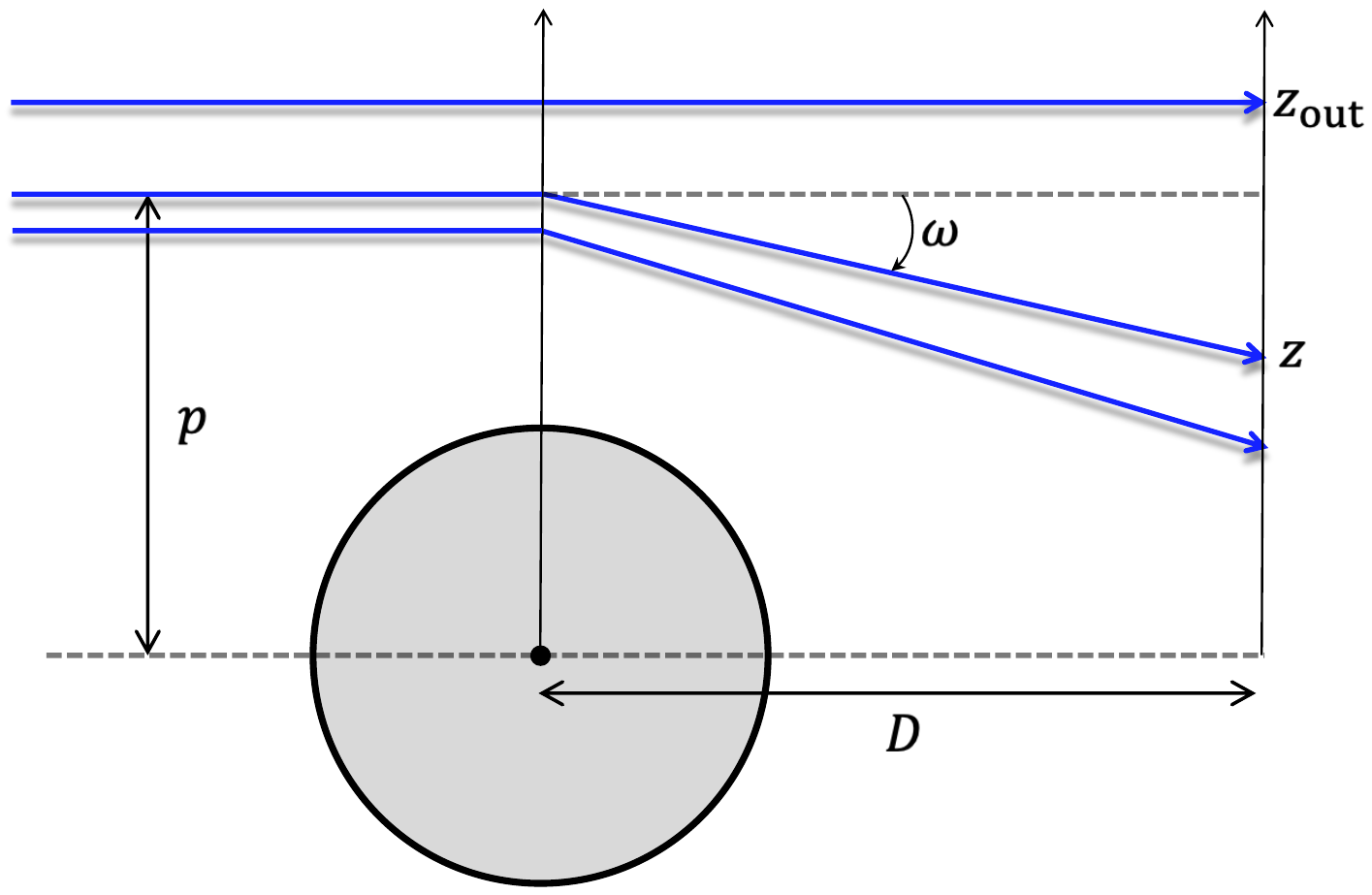}
\includegraphics[totalheight=4cm,trim=0 50 0 50]{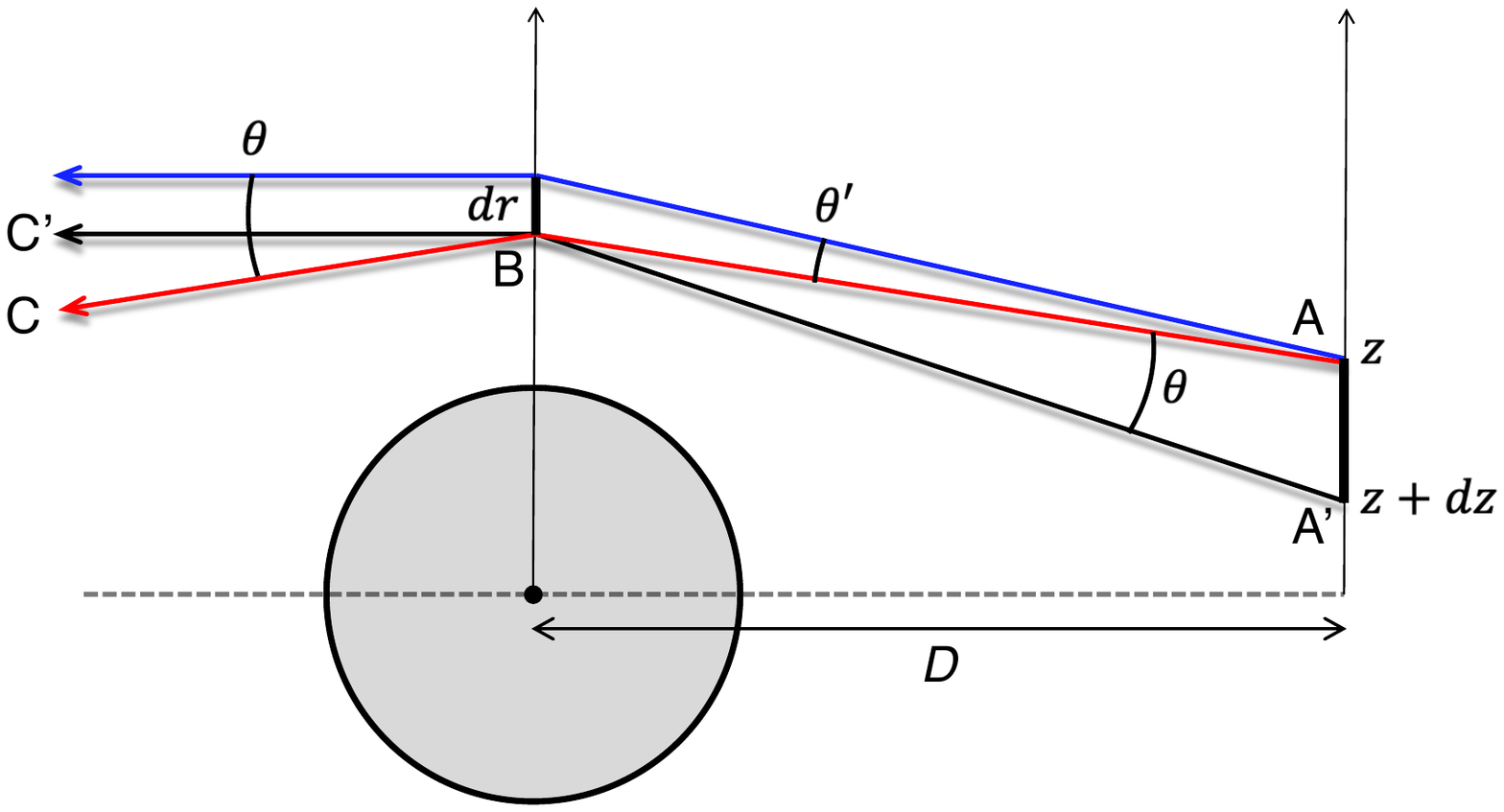}
\caption{%
\textit{Left:}
The geometry of a refractive occultation. 
The stellar rays come from infinity at left with impact parameter $p$. 
They are differentially refracted as they probe deeper atmospheric layers,
and reach the observer located at $z$ after traveling the distance $D$.
Outside the atmosphere (e.g. at $z_{\rm out}$), the deviation $\omega$ is zero. 
\textit{Right:}
A sketch illustrating the relation between 
the angular diameter $\theta$ of a star and the angular diameter of its refracted image, $\theta'$.
See text for details.
}%
\label{fig_dp_dz}
\end{figure}

Figure~\ref{fig_dp_dz} summarizes the principle of differential refraction in an atmosphere,
and defines the various geometrical quantities used in the text.
For the commodity of plotting, the deviation is sketched as an abrupt change of propagation, 
which it is actually gradual (Fig.~\ref{fig_deviation_total}).

%%%%%%%%%%%%%%%%%%%%%%%%%%%%%%%%%%%%%%%%%%
%%%%%%%%%%%%%%%%%%%%%%%%%%%%%%%%%%%%%%%%%%

\subsection{Abel inversion}

Stellar occultations provide the deviation $\omega$ of the ray and 
the corresponding impact parameter $p$ at the various instants of the event,
thus yielding $p(\omega)$.
The problem is now to derive the value of $n(r_0)$ using all these values $p(\omega)$.
This is done through an \it Abel inversion \rm of Eq.~\ref{eq_total_devia},
which is detailed for instance in \cite{fje71,vap73,was73} and provides
\begin{equation}
n_0= 
\exp \left\{ 
\frac{1}{\pi} \int_{\omega=0}^{\omega(p_0)}
\ln \left[ 
\frac{p(\omega)}{p_0} +\sqrt{ \left( \frac{p(\omega)}{p_0} \right)^2 -1}
\right] d\omega
\right\},
\label{eq_abel}
\end{equation}
which is a relation between $n_0$ and $p_0 = r_0 n_0$, from which $n_0$ corresponding to the distance $r_0$, 
i.e. $n(r0)$, is retrieved.

During an occultation by a transparent planetary atmosphere, the flux of the star gradually dims
due to the \it differential deviation \rm of the stellar rays. 
Thus in this case, the dimming of the flux is not caused by absorption or scattering (due for instance to hazes) or 
but by refraction. 

Consider in  Fig.~\ref{fig_dp_dz} a planet with an atmosphere that deviates a stellar ray
with impact parameter $p$ 
by an angle $\omega$ (which is negative by convention, see Figs.~\ref{fig_deviation_total} and \ref{fig_dp_dz}),
and reaches the observer at $z$. We have
\begin{equation}
z= p + D\omega,
\label{eq_p_z}
\end{equation}
which provides
$$
d\omega = \frac{1-(dp/dz)}{D}dz.
$$
If the atmosphere is transparent, the luminous flux contained in the beam of width $dp$ is retrieved
in the beam of width $dz$. 
Consequently, the stellar flux at $z$ is ``diluted" and yields the 
irradiance\footnote{We ignore for moment the curvature of the planetary limb. It becomes relevant near
the shadow center, where it causes a ``flash", see Section~\ref{sec_central_flash}.}:
\begin{equation}
\phi= \frac{dp}{dz},
\label{eq_dp_dz}
\end{equation}
taking a stellar irradiance outside the occultation normalized to unity.
Thus we obtain $\omega(z)$ by
\begin{equation}
\omega(z)=
\frac{1}{D} \int_{+\infty}^z (1-\phi) dz.
\label{eq_omega}
\end{equation}
In practice, it is enough to start the integration just outside the occultation,
at some level $z_{\rm out}$ where we have $\phi=1$ and $\omega=0$, see Fig.~\ref{eq_omega}.
This is reached rapidly, as planetary atmospheres decay exponentially in density,
and become undetectable by the observer above a certain level. 
Once $\omega(z)$ is known, $p(z)$ is given by Eq.~\ref{eq_p_z}, which provides $\omega(p)$.
This is finally introduced in Eq.~\ref{eq_abel} to obtain the refractivity profile $n_0$ at radius $r_0$.

Here we drop for sake of simplicity the index 0 and we use $r$ instead of $r_0$.
The refractivity of the gas at a given radius is defined as $\nu(r)= n(r)-1$. 
It is related to the molecular density of the gas $n_{\rm g}$ by
$\nu = K n_{\rm g}$,
where $K$ is the molecular refractivity of the gas under consideration.
Consequently, the Abel inversion eventually provides the density profile of the atmosphere through 
\begin{equation}
n_{\rm g}(r) = \frac{\nu(r)}{K}.
\label{eq_nu_ng}
\end{equation}
On the other hand, the hydrostatic equation provides the pressure $p$ by integration of the equation
\begin{equation}
\frac{dp}{dr}= -\mu n_{\rm g}(r) g(r),
\label{eq_p_r}
\end{equation}
where $\mu$ is the molecular mass and $g(r)$ the acceleration of gravity at radius $r$.
Finally, the ideal gas equation $p= n_{\rm g} k_B T$ (where $k_B$ is Boltzmann's constant) provides
\begin{equation}
\frac{1}{T} \frac{dT}{dr} =  
-\left[
\frac{\mu g(r)}{k_B T} +  \frac{1}{n_{\rm g}} \left( \frac{dn_{\rm g}}{dr} \right)
\right]
\label{eq_T_r}
\end{equation}

Both Eqs.~\ref{eq_p_r} and \ref{eq_T_r} are first order differential equations. 
As such, they require a boundary condition, 
i.e. the value of $p$ and $T$ at some prescribed radius $r$, respectively.
This is not too much of a problem for the pressure, as the atmosphere decays exponentially
with radius. Thus, we can safely take $p=0$ at $z_{\rm out}$, as we did for $\omega$.

This approximation cannot be used for the temperature, as it is usually \it not \rm known at $z_{\rm out}$,
and it is certainly \it not \rm zero.
In fact, integrating Eq.~\ref{eq_T_r} requires an independent knowledge of $T$ at some given radius. 
Otherwise, an infinity of profiles $T(r)$ can explain the same observable (here, the occultation light curve). 
This ambiguity can be resolved for instance by using other ground-based observations or
spacecraft measurements that have access to the level probed by the occultation.
Another approach is to propose physical arguments (such as a radiative transfer model) 
that restrict the range of plausible values of $T$ at some level.

\subsection{Conservation of energy, primary and secondary stellar images}

We now consider the problem of stellar images during a refractive occultation.
For this, we have to reverse the diagram displayed in the left panel of Fig.~\ref{fig_dp_dz}, 
as shown in the right panel of this same figure.
Let us consider an observer at coordinate $z$ (point $A$) who watches though the atmosphere a star at infinity 
with angular diameter $\theta$, subtended by the blue and red rays in the figure. 
These two rays are deflected by the atmosphere and reach the observer at $A$, where they subtend an angle $\theta'$.
This angle defines the angular diameter of the stellar image after refraction.
We now rotate the red ray around $B$ by an angle $\theta$, $A$ and $C$ will superimpose onto $A'$ and $C'$, respectively.
In that case, $BC'$ is parallel to the outgoing blue ray.

For large distances $D$, $\theta= dz/D$ and $\theta'= dr/D$. Using Fig.~\ref{fig_dp_dz} 
and $p \sim r$, we have from the conservation of energy in a transparent atmosphere
\begin{equation}
\frac{\theta'}{\theta}= \phi.
\label{eq_thetap_over_theta}
\end{equation}

This means that the stellar image is compressed by a factor of $\phi$ perpendicularly to the limb of the planet.
Consequently, the brightness of the stellar image through the atmosphere (i.e. the flux received per unit surface and unit angle
at the observer) is the same as the brightness of the stellar image outside the atmosphere. 
The equation above actually states the conservation of the specific intensity (also called radiance or brightness) 
of the ray as it propagates through the transparent atmosphere, a theorem due to Clausius 
%{\color{blue}
(see \cite{cla64} and the discussion in \cite{goo95} p. 1.25).
%}
%
Another, equivalent way to state this conservation law is to say that the observed irradiance is proportional to
the angular dimension of the stellar image seen through the atmosphere.

For the moment, we have considered that the refraction acts in the plane of the figure.
In reality, the limb curvature also causes a deviation of the rays, perpendicular to that limb.
It is then easy to extend the result obtained above to a full 2D image, replacing 
the angle $\theta$ (resp. $\theta'$) by the solid angle $\Omega$ (resp. $\Omega'$) subtended by the star (resp. its image). 
Then Eq.~\ref{eq_thetap_over_theta} can be re-written as
\begin{equation}
\frac{\Omega'}{\Omega}= \phi,
\label{eq_Omegap_over_Omega}
\end{equation}
which states again that the received flux during an occultation by a transparent atmosphere
is directly proportional to the apparent size of the refracted stellar image.

\section{Useful approximations}

\subsection{The straight line approximation}

\begin{figure}[!h]
\centering
\includegraphics[totalheight=4cm,trim=0 0 0 0]{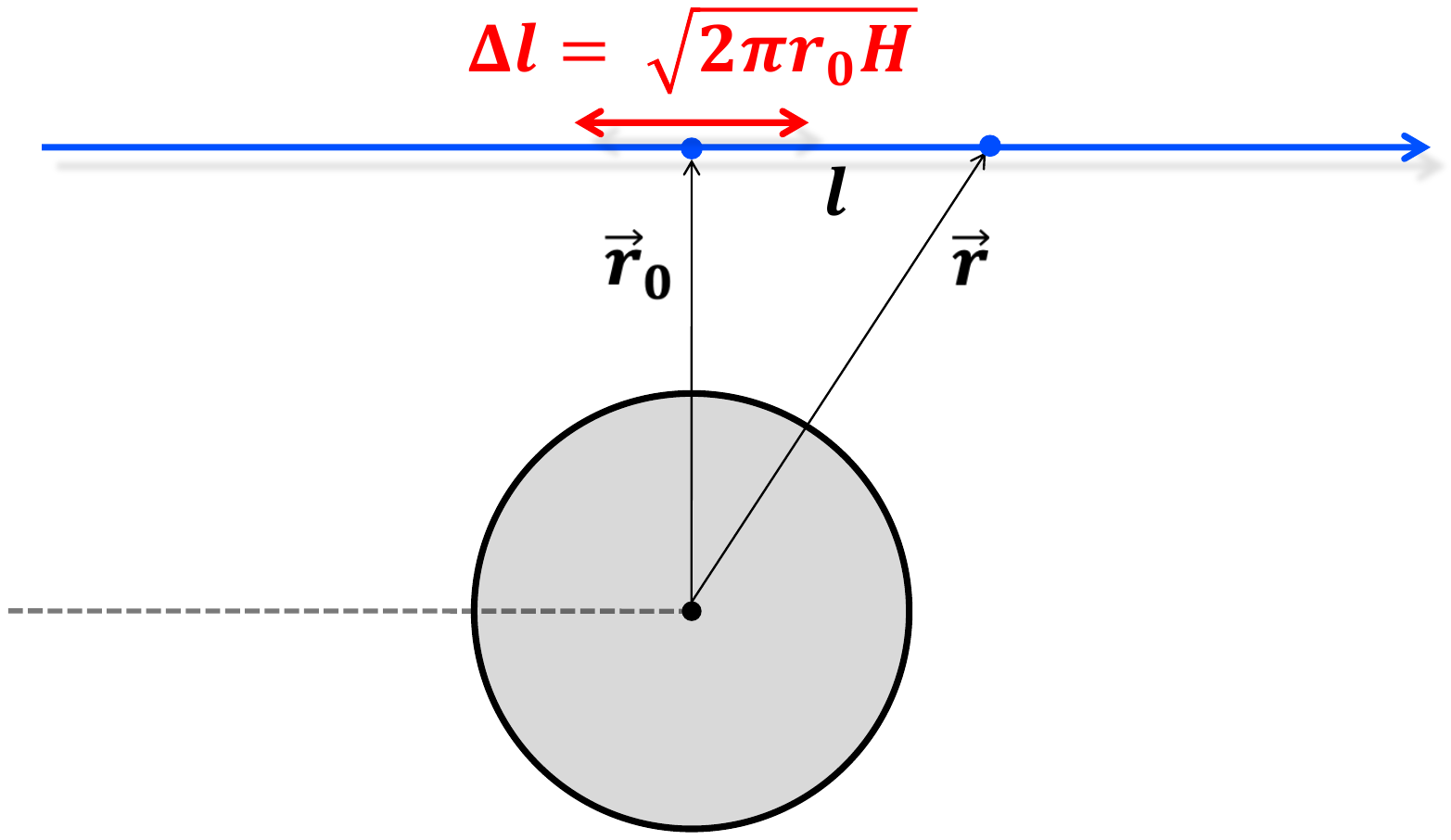}
\caption{%
The straight line approximation for refractive occultations by remote bodies in the solar system
with radius $r_0$ and atmospheric scale height $H$.
The deviation angle of $\omega (r_0)$ in Eq.~\ref{eq_straight_line} is so small that it is not perceptible at the scale of the figure.
The length $\Delta l$ (Eq.~\ref{eq_Delta_l}) corresponds to the interval where most of the ray deviation occurs.
}%
\label{fig_straight_line}
\end{figure}

For ground-based stellar occultations, 
the deflection angle $\omega$ (Eq.~\ref{eq_omega}) is very small because the distance $D$ is very large.
This angle is actually of the order of the angular diameter of the observed body, typically a few 
arc seconds, i.e. less than $10^{-5}$ radian.
From Eq.~\ref{eq_bouguer_rule}, this implies that $r_0$ is very close to $p_0$, 
to within $p_0 \omega$, so that $n_0$ is very close to unity.
Consequently, in Eq.~\ref{eq_total_devia}, we can write $n \sim 1$ and $p_0 \sim r_0$. 
Moreover, $dn/dr= d\nu/dr$, so that
$$
\omega (r_0) \sim  2 \int_{r_0}^{+\infty} r_0 \frac{d\nu}{dr}  \frac{dr}{\sqrt{r^2 - r_0^2}},
$$
One can change the variable $r$ to $l= \sqrt{r^2 - r_0^2}$ (Fig.~\ref{fig_straight_line}), which leads to
\begin{equation}
\omega (r_0) \sim  \int_{-\infty}^{+\infty} \left( \frac{r_0}{r} \right) \left( \frac{d\nu}{dr}  \right) dl,
\label{eq_straight_line}
\end{equation}
where $r= \sqrt{r_0^2 + l^2}$.
Note that the equation above can be used for ray tracing purposes, once a density profile $n_{\rm g}(r)$ 
-- and thus a refractivity profiles $\nu(r)$, see Eq.~\ref{eq_nu_ng}) -- has been prescribed.

%%%%%%%%%%%%%%%%%%%%%%%%%%%%%%%%%%%%%%%%%%
%%%%%%%%%%%%%%%%%%%%%%%%%%%%%%%%%%%%%%%%%%

\subsection{The small scale height approximation}

In most of the cases, planets have atmospheres with a roughly constant \it  density \rm scale height $H$,
defined as
$$
H= -\frac{n_g}{\left(dn_g/dr\right)}.
$$
In usual cases, $T$ varies much more slowly than $n_g$, so that $H$
also varies slowly with $r$. Moreover, it is  usually much smaller than the planet typical radius, i.e. 
\begin{equation}
H \ll r.
\label{eq_H_ll_r}
\end{equation}
Then
$$
\nu(r) \sim \nu_0 \exp \left[ -\left(\frac{r-r_0}{H} \right) \right] {\rm \ \ \ and\ \ \ }
\frac{d\nu}{dr} \sim -\left( \frac{\nu_0}{H} \right) \exp \left[ -\left(\frac{r-r_0}{H} \right) \right],
$$
where $\nu_0 = \nu(r_0)$.
Furthermore, we will see that most of the refractive deviation occurs over a distance 
$\Delta l$ that is significantly smaller than  $r_0$. 
Thus, 
$r= \sqrt{r_0^2 + l^2} = r_0 \sqrt{1+ l^2/l^2} \sim r_0 (1+ l^2/2r_0^2)$, 
so that $r-r_0 \sim  l^2/2r_0$ and $r_0/r \sim 1$. 
Using those approximations, 
introducing in Eq.~\ref{eq_straight_line} the expression of $d\nu/dr$ obtained above and
using $\int_{-\infty}^{+\infty} \exp(-u^2) du= \pi$, we obtain
\begin{equation}
\omega(r_0) \sim 
-\nu_0 \sqrt{\frac{2\pi r_0}{H}}.
% \sim \left( \frac{d\nu}{dr} \right)_0 \sqrt{2\pi r_0 H}.
\label{eq_devia_small_H}
\end{equation}
By writing $(d\nu/dr)_{r_0}= -\nu_0/H$, we obtain $\omega(r_0) \sim (d\nu/dr)_{r_0} \sqrt{2\pi r_0 H}$.
This shows that the deviation angle mainly comes from a interval along the ray path of characteristic length
\begin{equation}
\Delta l \sim \sqrt{2\pi r_0 H}.
\label{eq_Delta_l}
\end{equation}
For Jupiter, $r_0 \sim 70,000$~km and $H \sim 30$~km, so that $\Delta l \sim 3,500$~km, which is significantly smaller
that $r_0$, as announced.
For Pluto or Triton, $r_0 \sim 1500$~km and $H \sim 20-50$~km we obtain $\Delta l \sim 400-700$~km.
The approximation $\Delta l \ll r_0$ is not so good in those cases, but Eq.~\ref{eq_devia_small_H} still captures the
correct orders of magnitude.

\subsection{The Baum and Code equation}

Baum and Code \cite{bau53} derived a simple equation that describes how the stellar flux decrease when observed
from Earth when observing a stellar occultation by a planetary atmosphere.
Using Fig.~\ref{fig_dp_dz}, Eq.~\ref{eq_dp_dz} and $p \sim r$ (the straight line approximation), we obtain
\begin{equation}
\frac{1}{\phi} = 1 + D \left( \frac{d \omega}{d r} \right).
\label{eq_1_over_phi}
\end{equation}
We note that in Eq.~\ref{eq_devia_small_H}, the rapidly varying factor in an exponential atmosphere is $\nu_0$, not $r_0$.
More precisely, we have $d \omega/d r \sim -\omega/H$, so that, dropping from now on  the index 0 in Eq.~\ref{eq_devia_small_H}
\begin{equation}
\frac{1}{\phi} = 
1 - \frac{D\omega}{H}=
1 + \nu \sqrt{\frac{2\pi rD^2}{H^3}}.
\label{eq_1_over_phi_bis}
\end{equation}
This shows that the stellar flux has dropped by a factor of two (the ``half light level", denoted here by a subscript $1/2$) for 
\begin{equation}
\omega_{1/2}= -H/D, \ \ {\rm so \ that \ \ } z_{1/2}= r_{1/2} - H.
\label{eq_omega_1_2}
\end{equation}
This means that the ray corresponding to the half light level has been deviated by one scale height $H$ 
when it arrives at the observer. 
This occurs for 
\begin{equation}
\nu_{1/2}= \sqrt{\frac{H^3}{2\pi r_{1/2}D^2}},
\label{eq_nu_1_2}
\end{equation}
corresponding to a molecular density  $n_{\rm g, 1/2}= \nu_{1/2}/K$.

From Eqs.~\ref{eq_p_z} and \ref{eq_1_over_phi_bis} and 
from the definition of the half light level, we have
$$
\begin{array}{lllll}
 \displaystyle \frac{1}{\phi} =  &  \displaystyle 1 + \frac{r-z}{H}
& {\rm \ \ \ and\ \ \ } &
2  =                                        &  \displaystyle 1 +    \frac{r_{1/2}-z_{1/2}}{H}. \\
\end{array}
$$
Thus
\begin{equation}
\frac{1}{\phi} - 2 = \frac{r-r_{1/2}}{H} - \frac{\Delta z}{H},
\label{eq_1_over_phi_m_2}
\end{equation}
where 
$$
\Delta z= z - z_{1/2}.
$$ 
Using the fact that $\omega/\omega_{1/2} \sim \nu/\nu_{1/2} = \exp[-(r-r_{1/2})/H]$, 
and from Eq.~\ref{eq_omega_1_2}, we have
\begin{equation}
\omega = -\frac{H}{D} \exp \left[-\left(\frac{r-r_{1/2}}{H}\right)\right]
\label{eq_omega_vs_r}
\end{equation}
and thus from Eq.~\ref{eq_1_over_phi_bis}
\begin{equation}
\displaystyle
\frac{1}{\phi} - 1 = \exp \left[-\left(\frac{r-r_{1/2}}{H}\right)\right].
\label{eq_1_over_phi_ter}
\end{equation}
This permits to express $r-r_{1/2}$ as a function of $\phi$ in Eq.~\ref{eq_1_over_phi_m_2}, and finally get
\begin{equation}
\displaystyle
\left( \frac{1}{\phi} - 2 \right) + \ln \left( \frac{1}{\phi} - 1 \right) = -\frac{\Delta z}{H},
\label{eq_BC}
\end{equation}
known as the Baum and Code equation \cite{bau53}.

%%%%%%%%%%%%%%%%%%%%%%%%%%%%%%%%%%%%%%%%%%
\begin{figure}[!t]
\centering
\includegraphics[totalheight=5cm,trim=0 0 0 0]{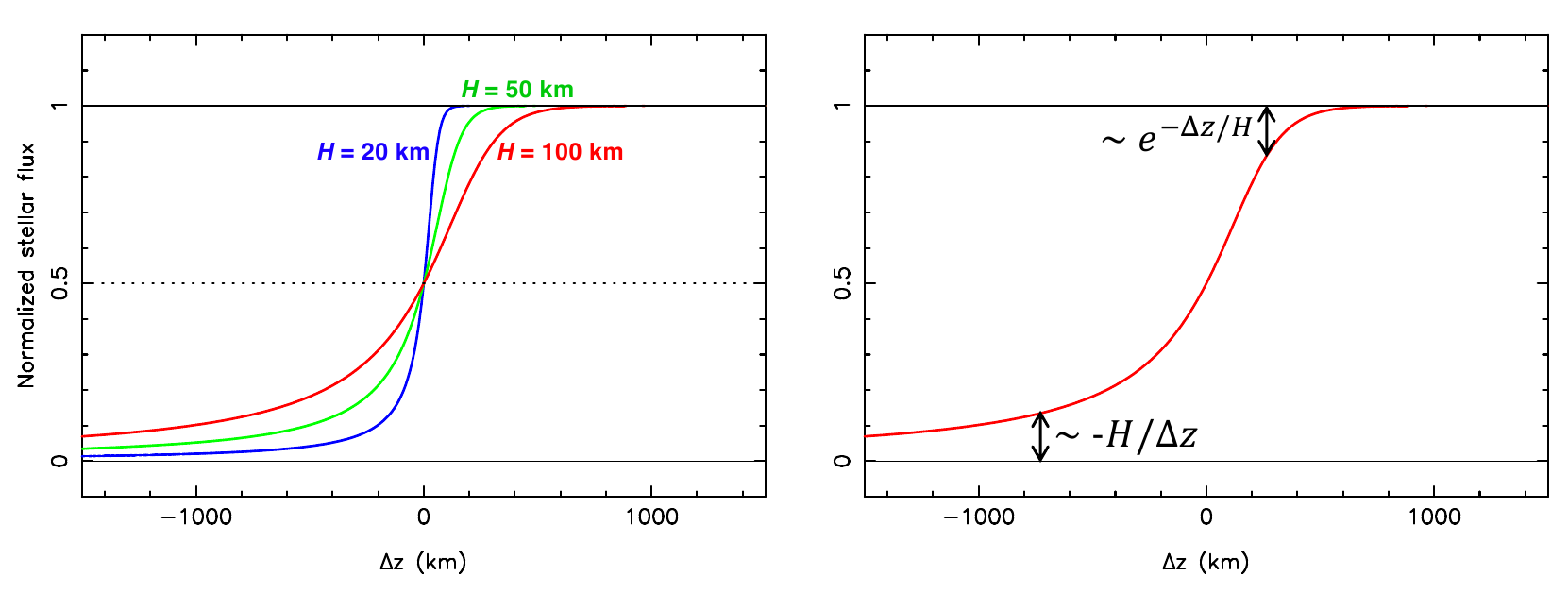}
\caption{%
\textit{Left:}
Examples of solutions to the Baum and Code equation~\ref{eq_BC} for typical values of the scale height $H$.
\textit{Right:}
The asymptotic behavior of the Baum and Code solution for large values of $\Delta z$ (positive or negative),
see Eqs.~\ref{eq_approx_BC_large_Dz_positive} and \ref{eq_approx_BC_large_Dz_negative}.
}%
\label{fig_baum_code}
\end{figure}
%%%%%%%%%%%%%%%%%%%%%%%%%%%%%%%%%%%%%%%%%%

Classical numerical schemes can be used to invert this equation so that to provide 
$\phi$ as a function of $\Delta z$, the distance traveled by the observer in the shadow plane (Fig.~\ref{fig_baum_code}).
Once this is done, Eq.~\ref{eq_1_over_phi_m_2} % plus the fact that $z_{1/2}= r_{1/2} - H$ 
provides the radius of closest approach $r$ probed by the ray as a function of $z$:
\begin{equation}
r= z + H \left( \frac{1}{\phi}-1 \right).
\label{eq_r_vs_z}
\end{equation}

Asymptotic expressions of $\phi$ can be obtained for $\Delta z \rightarrow +\infty$, 
i.e. when the star is observed far away from the planet. 
Then $\phi$ approaches unity, so that Eq.~\ref{eq_BC} yields
\begin{equation}
\phi \sim 1 -  \exp \left(-\frac{\Delta z}{H}\right).
\label{eq_approx_BC_large_Dz_positive}
\end{equation}
Consequently, the light curve approaches very rapidly (in fact, exponentially) the unocculted stellar flux unity
as the star probes a few scale heights only above the half light radius.
To take an example, suppose that 
the photometric quality of the occultation light curve is such that a drop of flux by at least 1\% is necessary 
to be detected, a common situation in practical cases.
This means that we must have $\exp(-\Delta z/H) > 0.01$ for detecting the stellar drop.
In other words, we can probe only levels below $r_{1/2} - \ln(0.01)H \sim r_{1/2}  + 4.6H$.

At the other extreme, in the case $\Delta z \rightarrow -\infty$
(i.e. when the star probes deep layers of the atmosphere), % and more precisely, layers that are much below the half light level.
Eq.~\ref{eq_BC} provides 
\begin{equation}
\phi \sim - \frac{H}{\Delta z}.
\label{eq_approx_BC_large_Dz_negative}
\end{equation}

Thus, the stellar flux goes to zero rather mildly ($\propto 1/|\Delta z|$) when compared to its exponential behavior near $\phi \sim 1$.
The asymptotic behaviors of $\phi$ are summarized in the right panel of Fig.~\ref{fig_baum_code}. 

\subsection{Applications to planetary atmospheres}

Eq.~\ref{eq_nu_1_2} shows that the larger the scale height $H$, the denser the half light level probed in the atmosphere.
Conversely, the larger the distance $D$, the smaller $n_{\rm g, 1/2}$. 
This explains why ground-based observations, for which $D$ is very large (up to billions of kilometers), 
can probe very tenuous pressure levels and still cause significant stellar drops.
Using the ideal gas equation and the classical expression $H \sim k_B T/\mu g(r_{1/2})$, 
Eq.~\ref{eq_nu_1_2}  provides the expression of the pressure $P_{1/2}$ probed by the half light rays:
\begin{equation}
P_{1/2} \sim \frac{GM\mu}{KD} \sqrt{\frac{H^5}{2\pi r_{1/2}^5}},
\label{eq_p_1_2}
\end{equation}
where $M$ is the mass of the body and $G$ is the constant of gravitation.
Using this equation and the parameters listed in Table~\ref{tab_p_1_2}, 
we obtain order-of-magnitude estimations of $P_{1/2}$.

\begin{table}[!h]
\caption{%
Estimation of the half light pressure level $P_{1/2}$ probed during ground-based stellar occultations.
\label{tab_p_1_2}
}%
\setlength{\tabcolsep}{1mm}
\begin{tabular}{lllllllll}
\hline \hline
Object & $GM$ 			& $r_{1/2}$ 	& $D$ 	& $H$ 	& gas 		& $\mu$		& $K$ (10$^{-29}$ & $P_{1/2}$ \\
	   & (m$^3$ s$^{-2}$) 	& (km) 		& (ua)	& (km) 	& composition	& (10$^{-26}$ kg) & m$^3$ molecule$^{-1}$) 	& (Pa) \\
\hline
Jupiter    & $1.21 \times 10^{17}$  & 71840 &  4.2  & 25 & 90\% H$_2$+10\% H$_{\rm e}$ & 0.37 & 0.479 & 0.1 \\
Saturn    & $3.79 \times 10^{16}$  & 61000 &  8.5  & 55 & 90\% H$_2$+10\% H$_{\rm e}$ & 0.37 & 0.479 & 0.2 \\ 
Uranus   & $5.79 \times 10^{15}$  & 25900 & 18.2 & 65 & 90\% H$_2$+10\% H$_{\rm e}$ & 0.37 & 0.479 & 0.2 \\
Neptune & $6.84 \times 10^{15}$  & 25100 & 29.0 & 50 & 90\% H$_2$+10\% H$_{\rm e}$ & 0.37 & 0.479 & 0.1 \\
Titan      & $8.98 \times 10^{12}$  &   3070  &  8.5  & 45 & N$_2$					 & 4.7    & 1.11   & 0.3 \\
Triton     & $1.43 \times 10^{12}$  &   1440  & 29.0 & 25 & N$_2$					 & 4.7    &  1.11  & 0.02 \\
Pluto     & $8.70 \times 10^{11}$   &   1300  & 32.0 & 65 &N$_2$ 					 & 4.7    &  1.11  & 0.2 \\
\hline
\end{tabular}
\end{table}

In Table~\ref{tab_p_1_2}, we assume that the objects are observed from Earth near opposition, providing a heliocentric distance $D$
that is roughly the orbital radius of the object minus 1~au.
In spite of very large ranges of values for the masses, radii, molecular masses and distances, 
we see that the combination of these parameters eventually provides a rather narrow range for $P_{1/2}$, 
typically a fraction of a Pascal, corresponding a few $\mu$bar, using 1 Pa = 10~$\mu$bar.

%{\color{blue}
The analysis presented in this Section assumes an atmosphere with constant scale height $H$
and adopts approximations that permit the use of the Baum and Code equation.
In view of the increasing quality of the occultation light curves and 
the significant departure of certain density profiles from having a constant $H$, 
it is now customary to use ``brute force" ray tracing that numerically integrate Eqs.~\ref{eq_total_devia} and \ref{eq_T_r},
and then Eqs.~\ref{eq_p_z}-\ref{eq_omega}.
This allows one to generate synthetic light curves with any atmosphere profiles, and 
compare them with observations, even in the case of strong local variations of $H$. 
Moreover, ray tracing has the advantage to account for the limb curvature, which is not
necessarily circular, or not even smooth, as developed in the next Section.
%}

In any case, the stellar occultation method has been very unique and productive for studying planetary atmospheres.
Besides the historical cases evoked in Section~\ref{sec_refrac_plan_atmo}, and without being exhaustive,  
we may cite 
the structure and extinction of the Martian upper atmosphere \cite{ell77},
the study of waves in Uranus' stratosphere \cite{fre82}, 
the discovery of Pluto's atmosphere in the 1980's \cite{bro95,hub88,ell89},
its seasonal three-fold pressure increase between 1988 and 2020 \cite{ell03,sic03,sic21},
the wave forcing by solar-induced sublimation at Pluto's surface \cite{toi10,fre15},
the structure and evolution of Neptune's stratosphere \cite{roq94},
the structure, zonal wind regime and haze properties of Titan's stratosphere \cite{hub93,sic06}
and Triton's atmosphere \cite{mar22}.

An important point is the complementarity between those ground-based observations and space exploration.
They may access different regions of the studied atmospheres and thus, provide a synoptic description of
these atmospheres. 
Stellar, solar or radio occultations have been performed by various spacecraft.
As they are much closer to the body than terrestrial observers, they probe much deeper layers 
as the quantity $D$ in Eq.~\ref{eq_p_1_2} is much smaller.

For instance, 
while ground-based Titan occultations typically probe a few $\mu$bar to some 100 $\mu$bar pressure levels,
the solar occultations observed by the Cassini spacecraft could reach layers with pressure of more than 10 mbar \cite{bel09}.

\section{Central flashes}
\label{sec_central_flash}

\subsection{Primary and secondary images}

\begin{figure}[!h]
\centering
\includegraphics[totalheight=5cm]{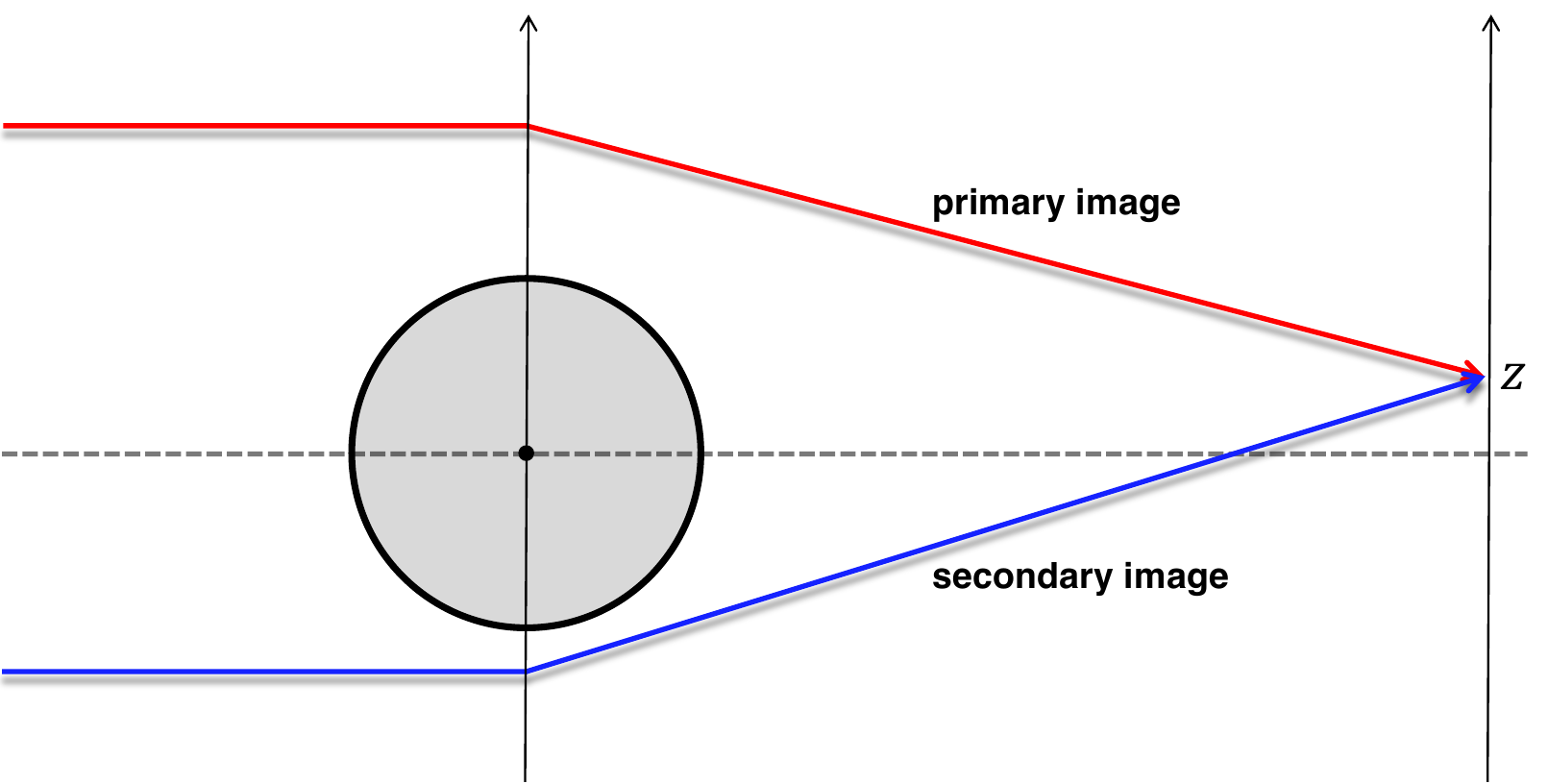}
\caption{%
Stellar rays coming from infinity at left can be refracted to the observed at $z$ following two paths. 
One produces the primary (or near-limb) image and corresponds to the less refracted ray, plotted here in red.
The other ray (in blue) produces a secondary (or far-limb) image that comes from the other side of the 
planetary disk, and thus suffers a stronger deviation.
If $z$ is negative, the primary and secondary characters of the images are swapped.
}%
\label{fig_prima_secon_rays}
\end{figure}

From now on, we will consider the curvature of the planetary limb, which creates a ``central flash" effect.
First, as illustrated in Fig.~\ref{fig_prima_secon_rays}, we see that a spherical planetary atmosphere generally produces two images,
a \textit{primary} image (sometimes called the near-limb image) and a \textit{secondary} (or far-limb) image. 
More complex situations where several images are produced by non-spherical atmospheres will be considered later in this chapter.

%%%%%%%%%%%%%%%%%%%%%%%%%%%%%%%%%%%%%%%%%%
%%%%%%%%%%%%%%%%%%%%%%%%%%%%%%%%%%%%%%%%%%

%%%%%%%%%%%%%%%%%%%%%%%%%%%%%%%%%%%%%%%%%%
\begin{figure}[!h]
\centering
\includegraphics[totalheight=5cm,trim=0 0 0 0]{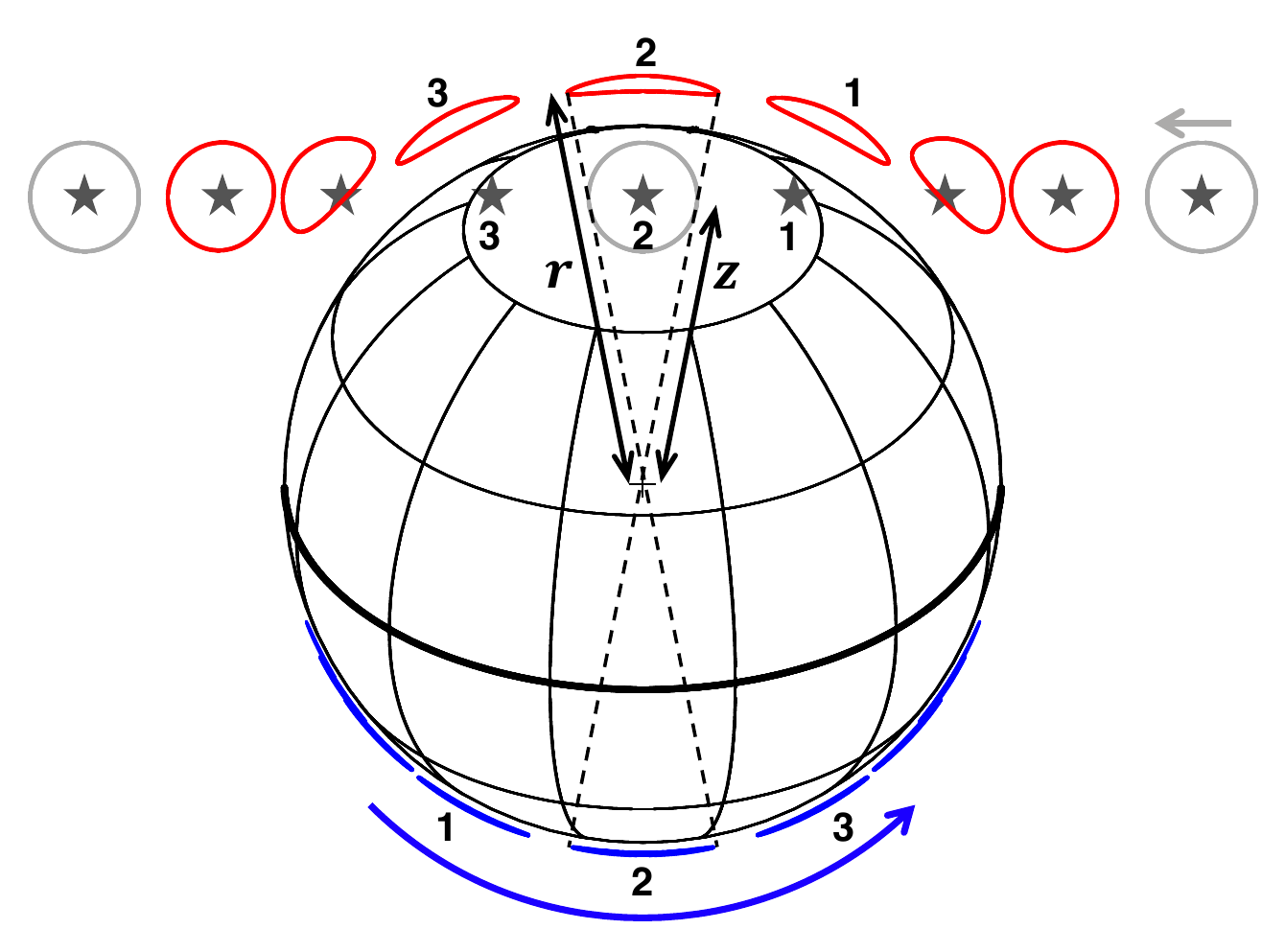}
\caption{%
The stellar images caused by refraction during a stellar occultation by an atmosphere.
Here, we consider a planet with radius $\sim$1200~km and an atmosphere with a scale height $H \sim 50$~km.
The gray circles with a star symbol at their centers delineate the apparent stellar disk projected at the planet distance, 
here with a radius $r_\star= 180$~km.
The plus symbol marks the planet center, assumed here to be spherical.
The occultation proceeds as the star moves from right to left relative to the planet.
The value of $r_\star$ has been greatly exaggerated here in order to illustrate the mechanism at play.
In real cases, $r_\star$ is at most a few kilometers when considering occultations by solar system objects.  
At any moment (for instance at points 1, 2 and 3), the stellar disk has two images. 
One is the primary (or near-limb, in red) image caused by the refraction due to the nearest point of the limb.
The other is the secondary (or far-limb, in blue) image caused by the opposite point of the limb,
i.e. rays that pass by the other side of the planet before reaching the observer.
The stellar images are compressed perpendicular to the limb due to differential refraction (Eq.~\ref{eq_thetap_over_theta}), and
they are stretched parallel to the limb due to the focusing caused by the limb curvature (Eq.~\ref{eq_phi_focusing}).
The stellar flux caused by a particular image is then proportional to the area encircled in the stellar image.
This latter can be calculated using the dashed lines. They show that the stretching greatly 
increases as the star approaches the planet center, as projected in the sky plane.
This leads to the detection of a central flash, see Fig.~\ref{fig_prima_secon_flash}.
}%
\label{fig_prima_secon_grazing}
\end{figure}
%%%%%%%%%%%%%%%%%%%%%%%%%%%%%%%%%%%%%%%%%%

The figure~\ref{fig_prima_secon_grazing} displays 
the compression and stretching suffered by the two stellar images in the spherical case,
which eventually explains the variation of flux observed during an occultation (Eq.~\ref{eq_Omegap_over_Omega}).
The compression of the image perpendicular to the limb (which decreases the received flux) 
is due to the differential refraction  that defocuses the stellar rays (Fig.~\ref{fig_dp_dz}) 
by a factor $1 + D (d \omega/d r)$ (Eq.~\ref{eq_1_over_phi}).
On the other hand, the stretching of the image along the limb (which increases the received flux) 
is caused by the limb curvature that focuses the stellar rays toward the shadow center by a factor $f$ 
(Fig.~\ref{fig_prima_secon_grazing}). Thus, the normalized irradiance received by the observer 
%{\color{blue}
from any of the two (or more) images produced by the limb is
%}
%
\begin{equation}
\phi = 
\left(  \frac{1}{\displaystyle 1 + D \frac{d \omega}{d r}}  \right) f \exp(-\tau) =
\left(  \frac{1}{\displaystyle 1 + D \frac{d \omega}{d r}}  \right) \left( \frac{r}{|z|} \right) \exp(-\tau).
\label{eq_phi_focusing}
\end{equation}
%
%{\color{blue}
The first equation above assumes that the stellar radius $r_\star$ projected at the planet 
distance\footnote{This is obviously different from the actual physical size of the star.}
is small compared to the atmospheric scale height $H$, so that the factor $dp/dz$ in Eq.~\ref{eq_dp_dz}
can be considered constant across the star diameter.
This is usually the case for planetary occultations, 
where $r_\star$ is of the order of a few kilometers and $H$ is of the order of a few tens of kilometers.
If $r_\star$ is comparable to or larger than $H$ (as it is the case in Fig.~\ref{fig_prima_secon_grazing}),
then from Clausius' theorem, $\phi$ is given by the surface area of the image normalized to $\pi r_\star^2$.
No analytical expression of this surface area is available in this case, but it can be easily obtained numerically.
%}

The second equation~\ref{eq_phi_focusing} is restricted to the case of a spherical atmosphere, 
for which $f= r/|z|$ from the examination of Fig.~\ref{fig_prima_secon_grazing}.
We have ignored this term so far because it is very close to unity near the half light level.
From Eq.~\ref{eq_omega_1_2}, $z_{1/2}= r_{1/2} - H$.
Thus at that level $f= r_{1/2}/|z_{1/2}|= r_{1/2}/(r_{1/2} - H) \sim 1$ because usually $H \ll r_{1/2}$.
In other words, the focusing effect due to the limb curvature is in general noticeable only near the shadow center.
%{\color{blue}
Morever, the second equation~\ref{eq_phi_focusing} is valid only if $|z|$ is larger than a few times $r_\star$.
We will see in the next subsection that the finite stellar size actually prevents the singularity that occurs at $|z|=0$
in Eq.~\ref{eq_phi_focusing}.
%}

We have also added in Eq.~\ref{eq_phi_focusing} a term $\exp(-\tau)$ which accounts for the presence of hazes that scatter and absorb
the light, where $\tau$ denotes the optical depth of the atmosphere along the line of sight. 
In several cases examined here, the atmosphere is transparent ($\tau=0$).
However, we will give examples on non-zero $\tau$, which leads to a decrease of flux
not only because of refraction, but also because of scattering and absorption.

%{\color{blue}
Once the irradiances of each image produced by the limb have been calculated using Eq.~\ref{eq_phi_focusing},
they are summed up to derive the synthetic light curve to be compared with the observations.
Note that in some cases, the planet is angularly large enough to distinguish the various stellar images moving along
the limb. Then, Eq.~\ref{eq_phi_focusing} can be used to produce synthetic light curves for each of these images.
%}

\subsection{Central flashes: the spherical case}

The equation~\ref{eq_phi_focusing} predicts that $\phi$ diverges to infinity at $z=0$, causing a ``central flash". 
This is true in the limiting case of a point-like source and geometric optics.
In actual cases, the star has a finite angular size, so that $f$ (and $\phi$) actually remains finite at 
the shadow center\footnote{However, even for a point-like source, the flux does not diverges at $z=0$ due to diffraction effects
that are not considered here.}.

The entire atmosphere can be seen as a lens that focuses the stellar rays toward the shadow center, that can be seen
as the focal point of that lens. 
More precisely, there is one layer (called the flash layer hereafter) in the atmosphere that has the right focal
length $D$, i.e. that causes the right deviation of the ray so that the observer can see the central flash.
Thus, the flash can be observed for any value of $D$, the flash layer being located deeper and deeper 
as $D$ decreases. 
Note that the flash ceases to be observed if the flash layer reaches the planet surface,
in which case the stellar image vanishes behind the limb.

%%%%%%%%%%%%%%%%%%%%%%%%%%%%%%%%%%%%%%%%%%
\begin{figure}[!h]
\centering
\includegraphics[totalheight=8cm,trim=0 0 0 0]{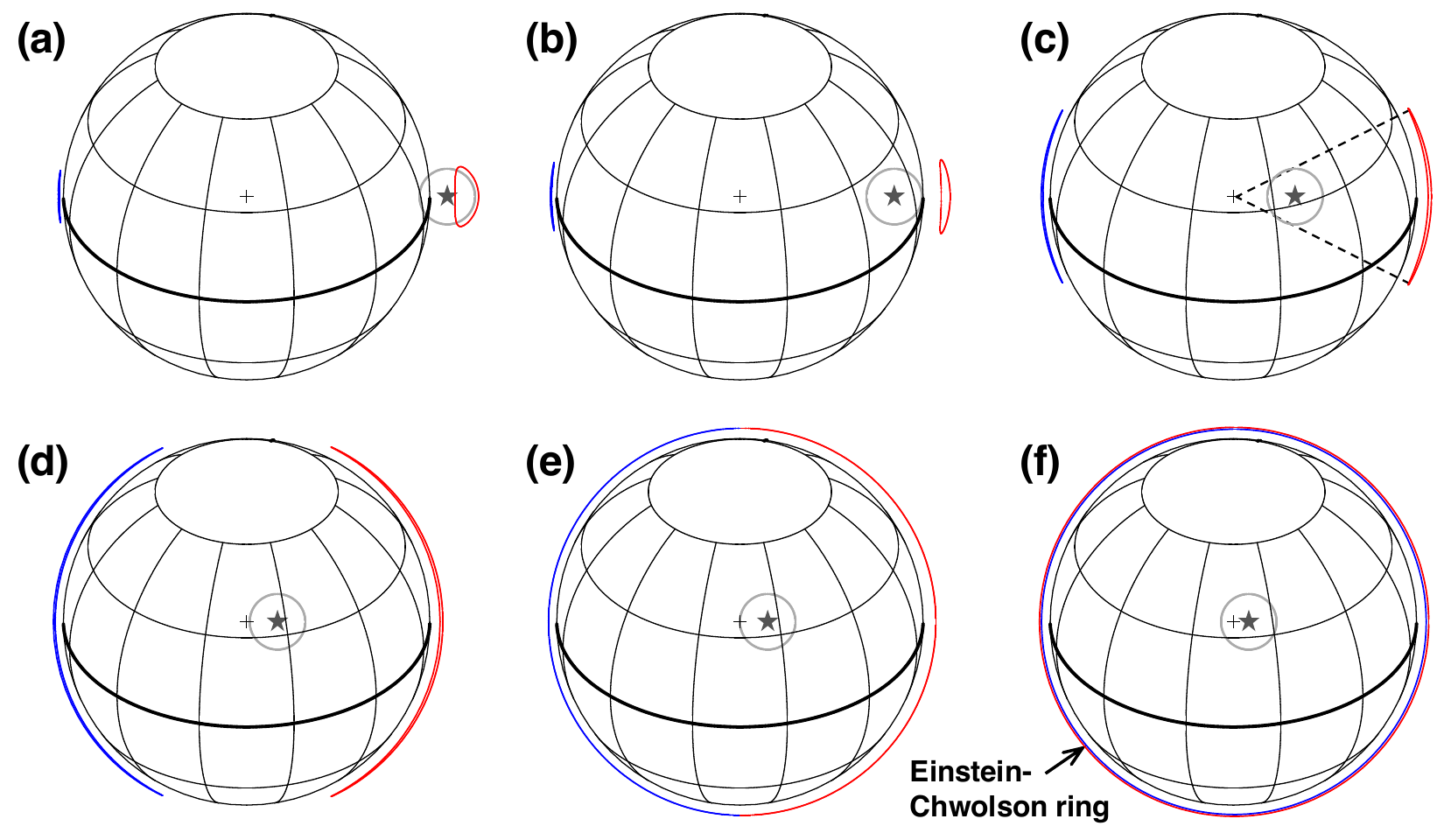}
\caption{%
The sequence leading to a central flash. 
This is the same as Fig.~\ref{fig_prima_secon_grazing}, but for a star that goes just behind the planet center (cross).
In steps (a)-(c), the primary image is first compressed parallel to the limb, 
and then stretched parallel to it due to, as illustrated by the dashed lines.
As the star aligns with the planet center, the primary and secondary images continue to stretch,
until they connect (step (e)) when the stellar limb intersects the planet center (i.e. the cross).
At this point, the two images merge into one and form a luminous ``Einstein-Chwolson ring" around the body,
%{\color{blue}
as illustrated in step (f).
%}
}%
\label{fig_prima_secon_flash}
\end{figure}
%%%%%%%%%%%%%%%%%%%%%%%%%%%%%%%%%%%%%%%%%%

The figure~\ref{fig_prima_secon_flash} summarizes the flash process. 
As long as the apparent stellar disk (delineated in gray in Fig.~\ref{fig_prima_secon_flash}) does not
overlap with the planet center (the plus symbol), the primary and secondary images are disconnected.
As soon as the stellar disk covers the center, these two images merge into a luminous ring surrounding the planet (panel (f)). 
This ring actually reveals the flash layer, i.e. the layer that focuses the stellar rays towards the shadow center.
This phenomenon is akin to the ``Einstein-Chwolson ring" caused by the gravitational lensing of rays coming from a remote 
galaxy or star by an intervening massive object.
So, although the causes of the bending are different (refraction vs. gravitation), 
atmospheric occultations and gravitational lenses share the same basic 
process\footnote{For massive planets like Jupiter, the ray bending caused by the gravity of the planet
is not negligible compared to the bending caused by refraction. However, its derivative $d \omega/d r$
is negligible and thus not affect significantly the stellar flux in Eq.~\ref{eq_phi_focusing}.}.

For a transparent atmosphere,
the brightness of that ring is the same as the brightness of the unocculted star 
(see the Clausius theorem discussed after Eq.~\ref{eq_thetap_over_theta}). 
so the flux received at perfect alignement star-planet-observer remains finite, 
contrarily to what is expected from Eq.~\ref{eq_phi_focusing}.
From Eq.~\ref{eq_thetap_over_theta} the width of the Einstein-Chwolson ring is
\begin{equation}
w_{EC} = 2r_\star \phi_{\rm c},
\label{eq_width_einstein_ring}
\end{equation}
where $\phi_{\rm c}$ is the stellar flux at the shadow center \it without \rm the focusing term $f$ in Eq.~\ref{eq_phi_focusing}.
Thus, the total area of the ring is $(2\pi r_{\rm cf}) w_{EC}$, where $r_{\rm cf}$ is the radius of the flash layer.
As the surface area of the star projected at the planet is $\pi r_\star^2$, the maximum height of the flash at shadow center is, 
normalized to the unocculted stellar flux and from Clausius theorem,
\begin{equation}
\phi_{\rm cf} = 
\frac{(2\pi r_{\rm cf}) w_{EC}}{\pi r_\star^2}= 
4 \left( \frac{r_{\rm cf}}{r_\star} \right) \phi_{\rm c},
\label{eq_phi_flash}
\end{equation}
where the second equations stems from Eq.~\ref{eq_width_einstein_ring}.
For order of magnitude considerations, we can use $\phi_{\rm c} \sim H/r_{\rm cf}$ (Eq.~\ref{eq_approx_BC_large_Dz_negative}), so that
\begin{equation}
\phi_{\rm cf} \sim 4 \left( \frac{H}{r_\star} \right).
\label{eq_phi_flash_approx}
\end{equation}
The height of the flash decreases as $r_\star$ increases because
the flash gets more and more convolved by the stellar disk. 
%Note also that $\phi_{\rm c}$ decreases with $H$, since the width of the ring $\w_{EC}$ is proportional to $\phi_{\rm c}$, and hence, to $H$.
Typical values of $H$ are $\sim$20-50~km depending on the planet, while $r_\star$ is typically of the order of one kilometer.
Thus, $\phi_{\rm cf}$ may reach values as large as one hundred or more at the very center of the shadow
of a spherical and transparent atmosphere.
This is indeed the case for Pluto and Triton's atmospheres, as seen later in this section. 

\subsection{Central flashes: the non spherical case}

If the atmosphere is not spherical, the equation~\ref{eq_phi_focusing} is still valid,
but the factor $|z|$ must be understood as the distance of the observer to the center of curvature of the flash layer.

\begin{figure}[!h]
\centering
\includegraphics[totalheight=5cm,trim=0 0 0 0]{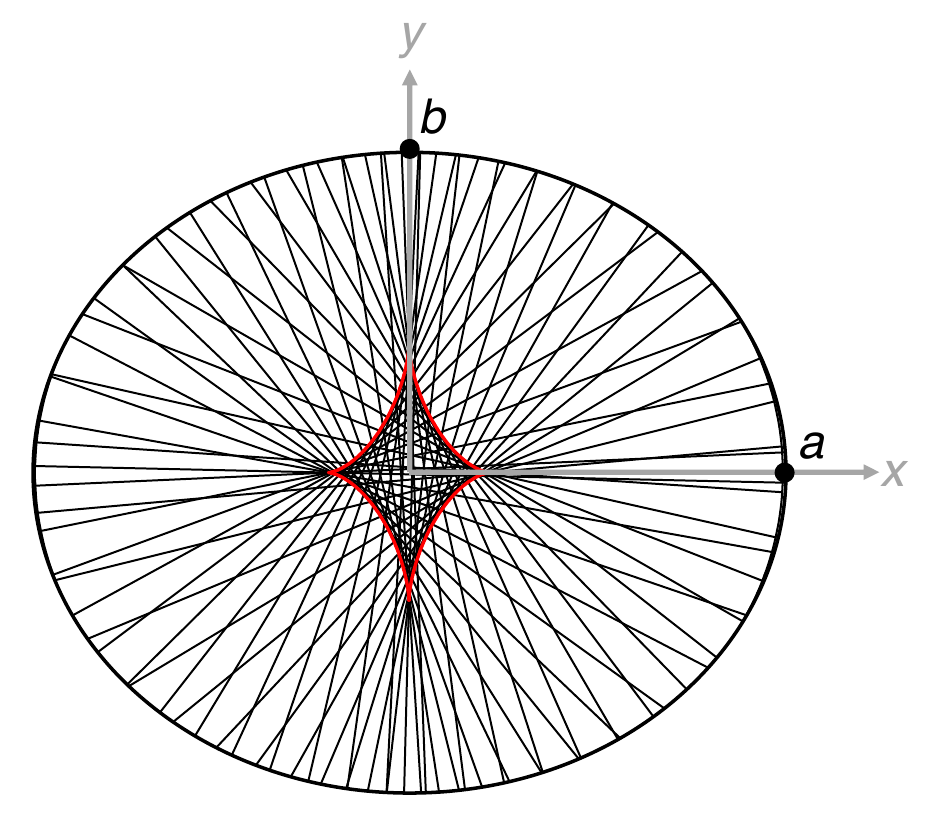}
\includegraphics[totalheight=5.15cm,trim=0 0 0 0]{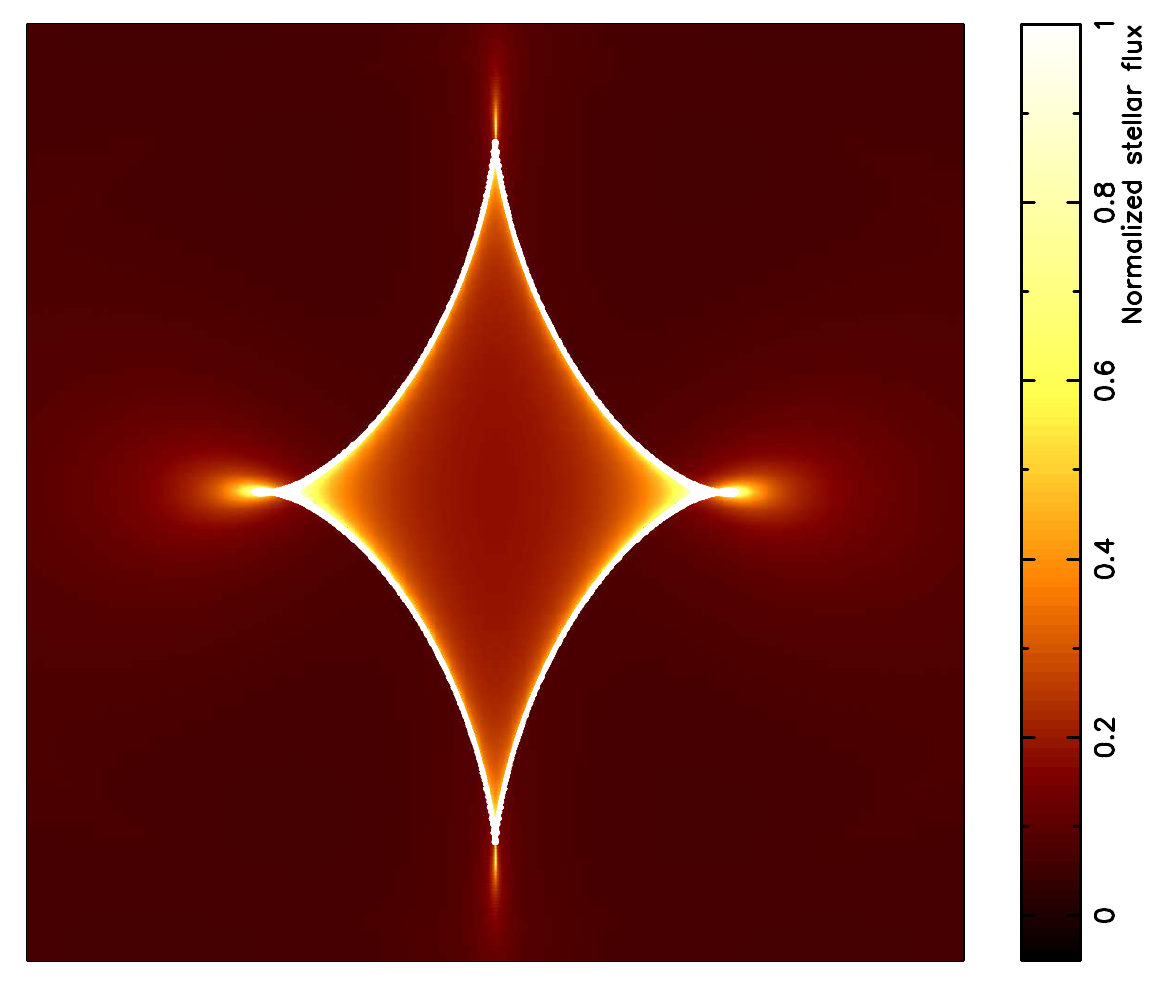}
\caption{%
\textit{Left:}  
a sketch showing the stellar ray deflected perpendicular to the limb of an oblate (here elliptical) atmosphere.
The rays converge towards the centers of curvature (or evolute) of each point of the limb.
Thus the stellar rays are everywhere tangent to the evolute.  
This creates a caustic (in red) where the observer detects discontinuous variations of flux. 
\textit{Right:} 
a close in view of the left panel, showing the intensity map of the flash near the shadow center.
Note the abrupt variation of flux along the caustic. 
}%
\label{fig_caustique_epsilon_0p1}
\end{figure}

A simple case is when the flash layer assumes a spheroid shape, i.e. an ellipsoid 
with equatorial and polar radii $a$ and $b$, respectively (Fig.~\ref{fig_caustique_epsilon_0p1}).
%
% Let us define the oblateness as $\epsilon= (a-b)/a$.
%This shape may be 
%{\color{blue}
due to the flattening of the solid planet itself (as it is the case for Mars \cite{ell77}),
%},
or may be maintained by zonal winds, i.e. an atmospheric flow parallel to the equator.
These winds create a centrifugal acceleration in a reference frame rotating with the planet.
It results in a flattening of the atmosphere under the combined effect of gravity.
Elliptical shapes have been used to describe central flashes observed during occultations by 
Mars \cite{ell77} or Neptune \cite{lel86}.

%%%%%%%%%%%%%%%%%%%%%%%%%%%%%%%%%%%%%%%%%%
%%%%%%%%%%%%%%%%%%%%%%%%%%%%%%%%%%%%%%%%%%

Each stellar ray is then deflected perpendicular to the limb of the planet, 
and converge towards the centers of curvature of the limb (called the evolute) to which the rays are tangent. 
This creates a caustic where the stellar flux suffers a sudden increase.
In the example of Fig.~\ref{fig_caustique_epsilon_0p1}, the flash layer appears with an elliptical shape whose equation is
$$
\left( \frac{x}{a} \right)^2 + \left( \frac{y}{b} \right)^2 = 1.
$$
The evolute of the ellipse has then the following equation (see e.g. \cite{bey76}),
$$
(a x)^{2/3} + (b y)^{2/3}= (a^2 - b^2)^{2/3},
$$
which is shown in red in Fig.~\ref{fig_caustique_epsilon_0p1}.

%%%%%%%%%%%%%%%%%%%%%%%%%%%%%%%%%%%%%%%%%%
\begin{figure}[!t]
\centering
\includegraphics[totalheight=4cm,trim=0 0 0 0]{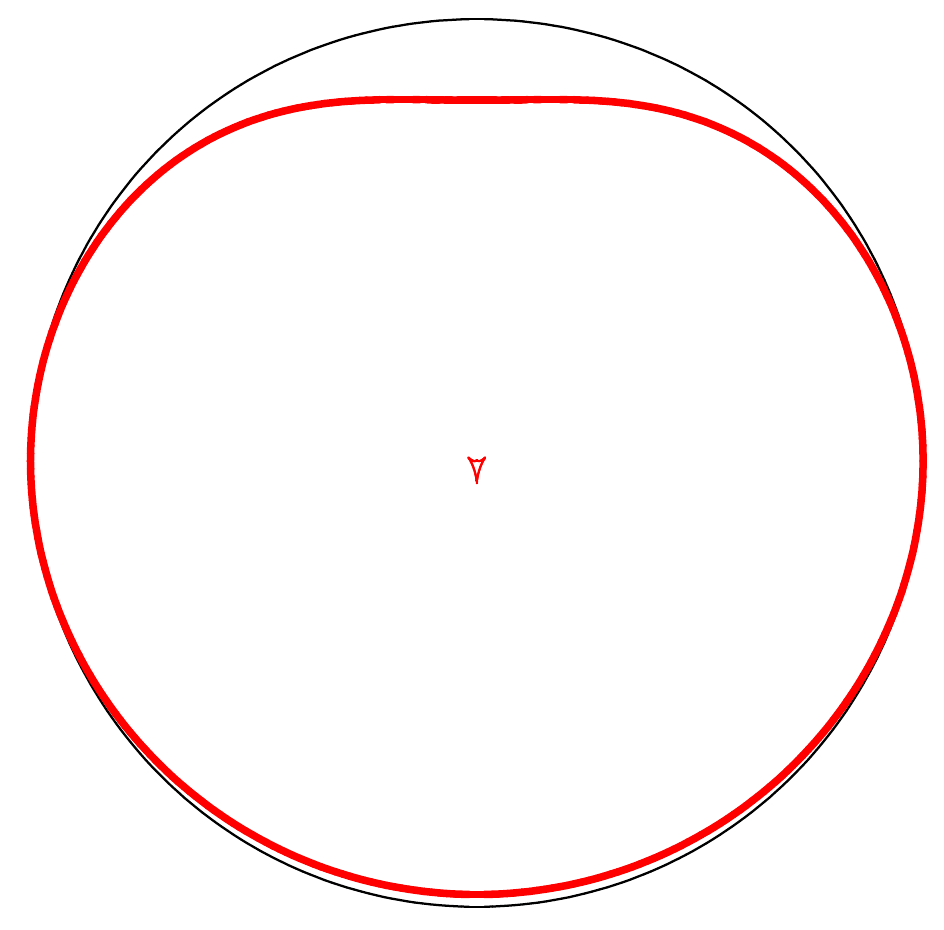}
\includegraphics[totalheight=4cm,trim=0 0 0 0]{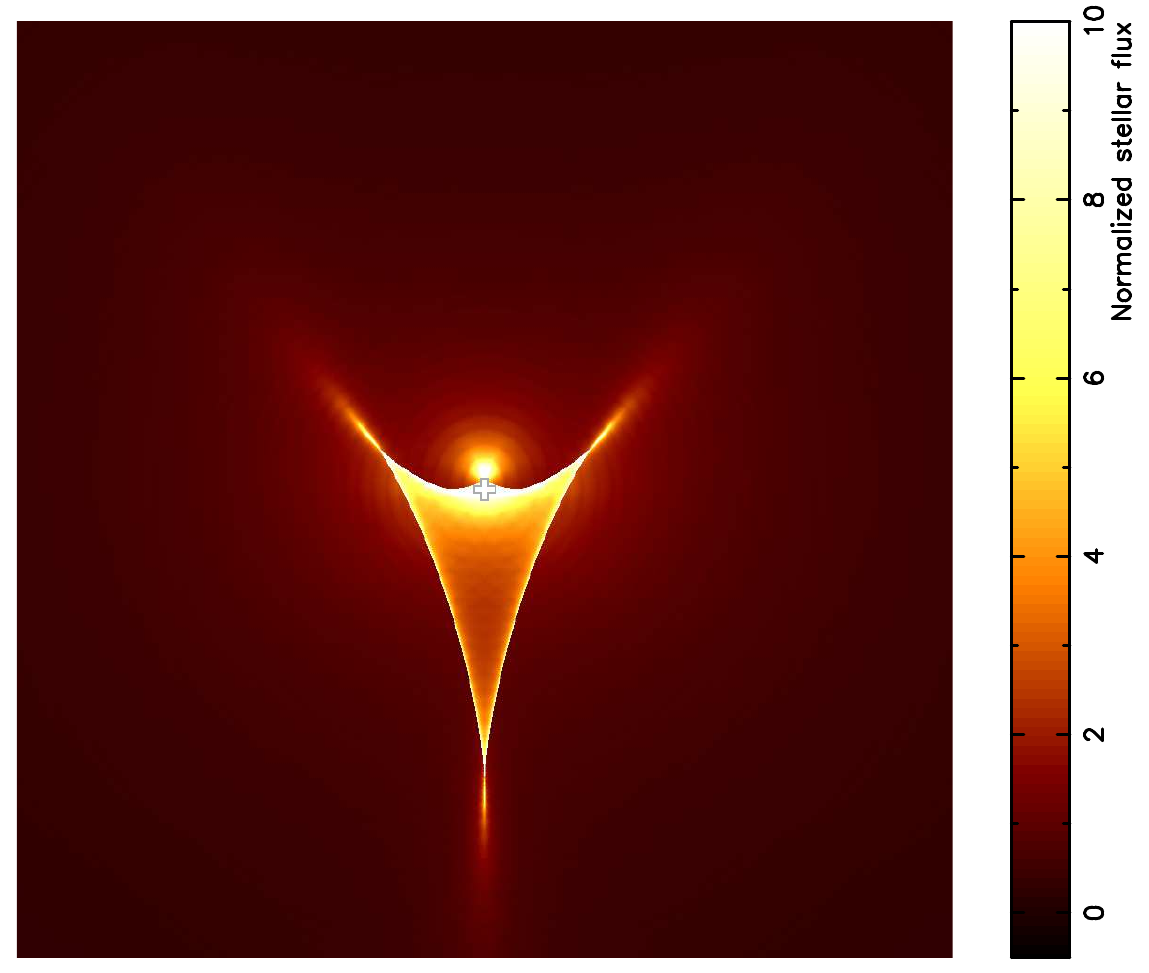}
\caption{%
\textit{Left:} 
Titan's central flash layer (as observed in November 2003) is delineated in black. 
It has a non-circular shape (red line) that is made more visible by expanding by a factor of twenty its departure from circularity.
While Titan's southern hemisphere is very close to spherical, 
its northern hemisphere is flattened by a strong zonal wind of about 200~m~s$^{-1}$ near latitude 60 deg north.
The limb evolute is plotted in red near the shadow center.
\textit{Right:} 
A close in view of the central flash, with Titan's center marked by a gray cross.
}%
\label{fig_caustique_Titan_03}
\end{figure}
%%%%%%%%%%%%%%%%%%%%%%%%%%%%%%%%%%%%%%%%%%

This said, there is no reason why the simple elliptical shape applies in all circumstances.
For instance, at its solstice, Titan has weak zonal wind regime in its summer hemisphere (which is then essentially spherical), 
and a strong jet in the winter hemisphere around the latitude 60 degrees \cite{hub93,sic06}.
The resulting shape of the atmosphere is delineated in red in the left panel of Fig.~\ref{fig_caustique_Titan_03},
with an expansion factor of twenty applied for a better viewing. 
The resulting intensity map (right panel) is then quite different from the elliptical case shown in Fig.~\ref{fig_caustique_epsilon_0p1}.

\begin{figure}[!h]
\centering
\includegraphics[totalheight=6cm,trim=0 0 0 0]{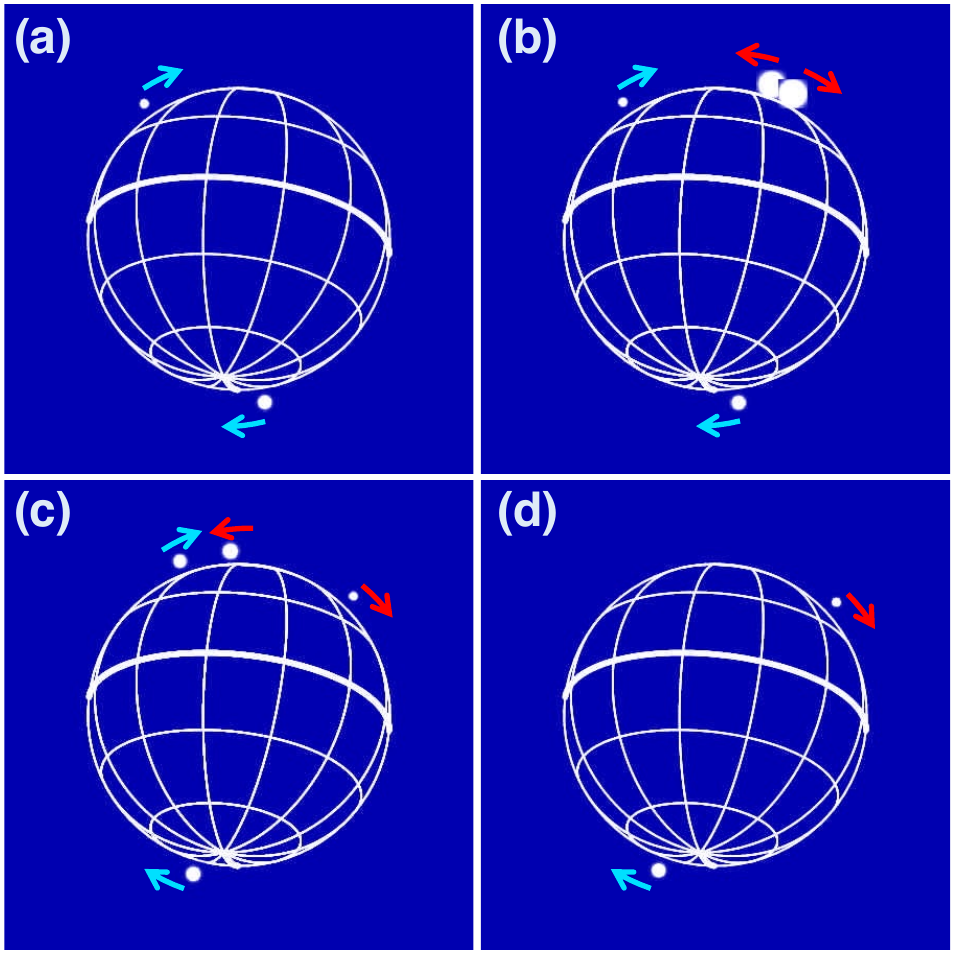}
\includegraphics[totalheight=6cm,trim=0 0 0 0]{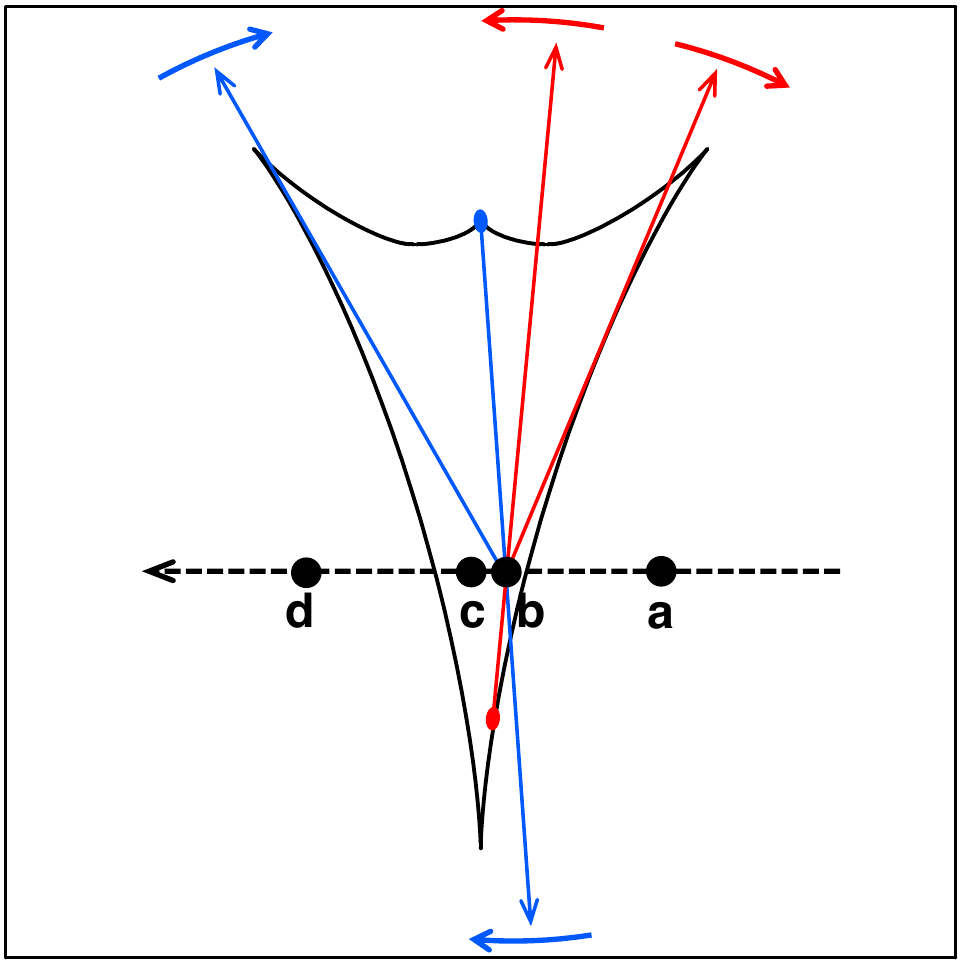}
\caption{%
\textit{Left:} 
the motion of the stellar images during an occultation by Titan, 
observed from Gifberg (Republic of South Africa) on 14 November 2008 \cite{sic06}.
These stellar images are \textit{reconstructed} from the observations of the flash. 
They could not be seen individually in the data, as Titan was too small (about one arcsec)
to be resolved by the instruments used during this campaign.
As long as the observer is outside the region delimited by the caustic (see right panel), 
only the two classical primary and secondary stellar images are detected.
When the observer is inside this region, four images contribute to the total flux.
They move rapidly along the limb,
eventually leading to the coalescence and disappearance of two of them as the observer leaves the caustic domain.
\textit{Right:} 
A close in view of the shadow center. 
It shows the position of the observer relative to the caustic, at each steps (a), (b), (c) and (d) illustrated at left. 
The straight arrows point to the stellar images seen by the observer at step (b). 
Those straight lines are the four solutions that pass through the point b, while being tangent to the caustic.
}%
\label{fig_gifberg_movie}
\end{figure}

An observer who is far away from the shadow center receives the flux from the two classical stellar images (primary and secondary)
moving in opposite directions (the blue arrows in panel (a) of Fig.~\ref{fig_gifberg_movie}).
The crossing of the caustic causes the sudden appearance of two bright stellar images that moves in opposite directions
(red arrows in panel (b)), so that four images are now seen.
As the observer proceeds towards the other side of the caustic, the two top-most images approaches each other
(blue and red arrows in panel (c)). They coalesce into a bright image at the crossing of the caustic, before disappearing suddenly.
As the observer recedes away from the caustic, only the two classical primary and secondary images remain (panel (d)).
The right panel of Fig.~\ref{fig_gifberg_movie}) illustrates how the positions of the four images can be
determined at any moment from the shape of the caustic in the case (b).

%%%%%%%%%%%%%%%%%%%%%%%%%%%%%%%%%%%%%%%%%%
%%%%%%%%%%%%%%%%%%%%%%%%%%%%%%%%%%%%%%%%%%

%%%%%%%%%%%%%%%%%%%%%%%%%%%%%%%%%%%%%%%%%%
\begin{figure}[!h]
\centering
\includegraphics[totalheight=4cm,trim=0 0 0 0]{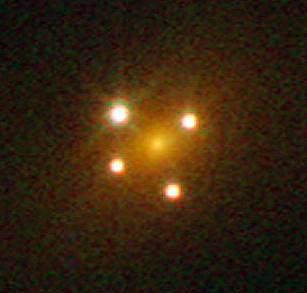}
\includegraphics[totalheight=4cm,trim=0 0 0 0]{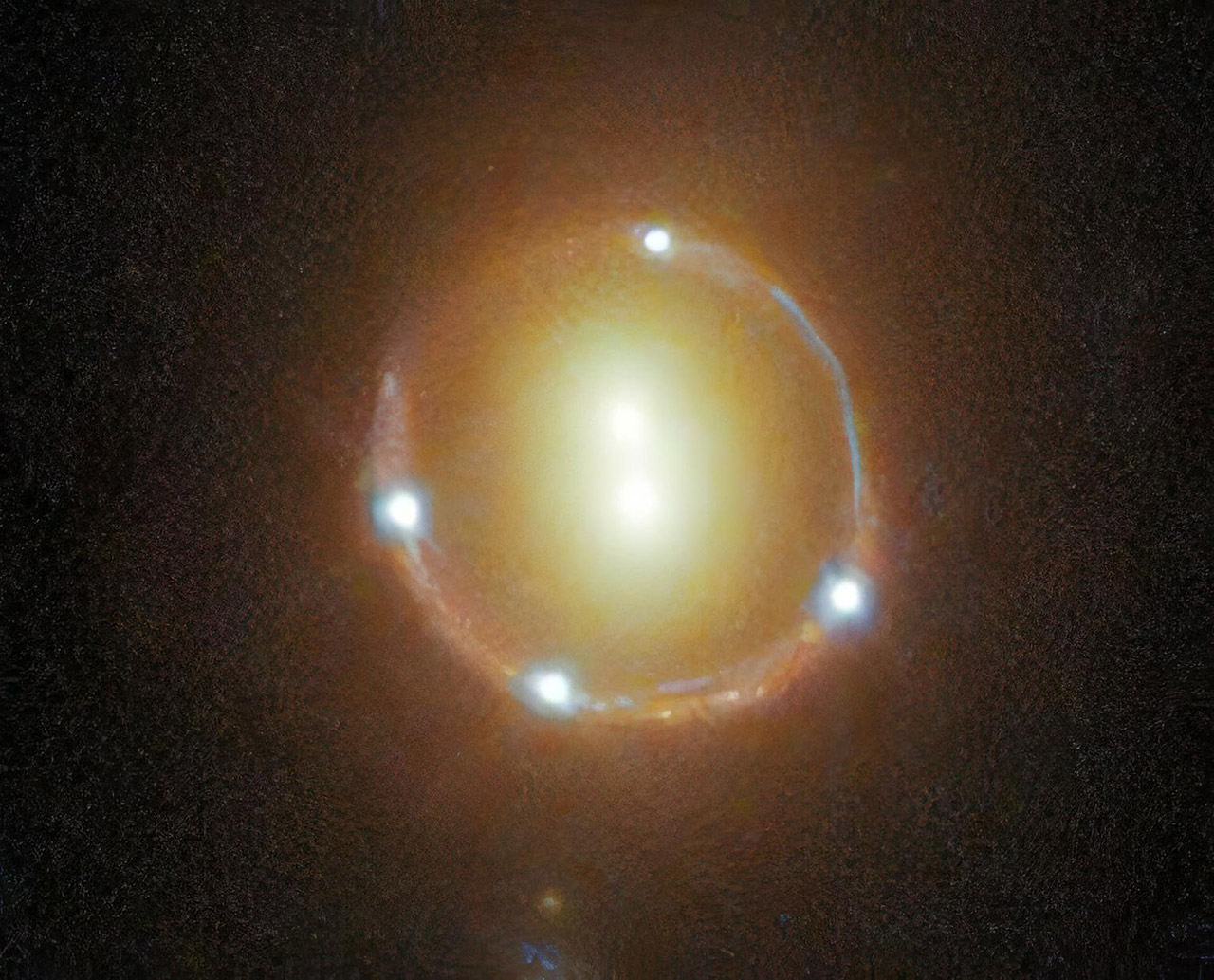}
\caption{%
\textit{Left:}
An example of a quadruple image (or Einstein cross) caused by a gravitational lensing 
of a remote quasar by the foreground galaxy UZC J224030.2 032131,
the diffuse object seen inside the four quasar images.
This is similar to the quadruple images seen in the images (b) and (c) Fig.~\ref{fig_gifberg_movie} (left panel).
\textit{Right:}
the same where the lens is now a double galaxy (the diffuse objects at the center).
Again an Einstein cross is visible, with four images of the quasar 2M1310-1714,
but also an Einstein-Chwolson ring, which is the image by the lens
of the extended galaxy which hosts the quasar. 
Credit: ESA/Hubble and NASA.
}%
\label{fig_einstein_ring}
\end{figure}
%%%%%%%%%%%%%%%%%%%%%%%%%%%%%%%%%%%%%%%%%%

\subsection{Einstein-Chwolson ring and Einstein cross}

In the case of stellar occultations by bodies such as Titan, Pluto or Triton, the angular resolution of classical
imaging is usually not sufficient to resolve the disk of these objects (that are at the level of one arcsec or less)
and thus see the stellar images moving along the limb.

As mentioned earlier, gravitational lenses act on objects like galaxies that are much more extended angularly
than the planetary bodies mentioned above. 
It is then possible to resolve the images and obtain a direct illustration of the Einstein-Chwolson ring. 
If the source is a point-like object (like a quasar), it is even possible to see the various images provided by the foreground lens,
for instance the four images seen in panels (b) and (c) of Fig.~\ref{fig_gifberg_movie}, often dumbed as the ``Einstein cross".
An example of Einstein cross is provided in the left panel of Fig.~\ref{fig_einstein_ring}.
The right panel displays an image where both the Einstein cross and the Einstein-Chwolson ring are seen.

%%%%%%%%%%%%%%%%%%%%%%%%%%%%%%%%%%%%%%%%%%
\begin{figure}[!b]
\centering
\includegraphics[totalheight=4cm,trim=0 0 0 0]{map_2D_gcm03_cuts_m0p5_10_Titan_orange.pdf}
\includegraphics[totalheight=4cm,trim=0 0 0 0]{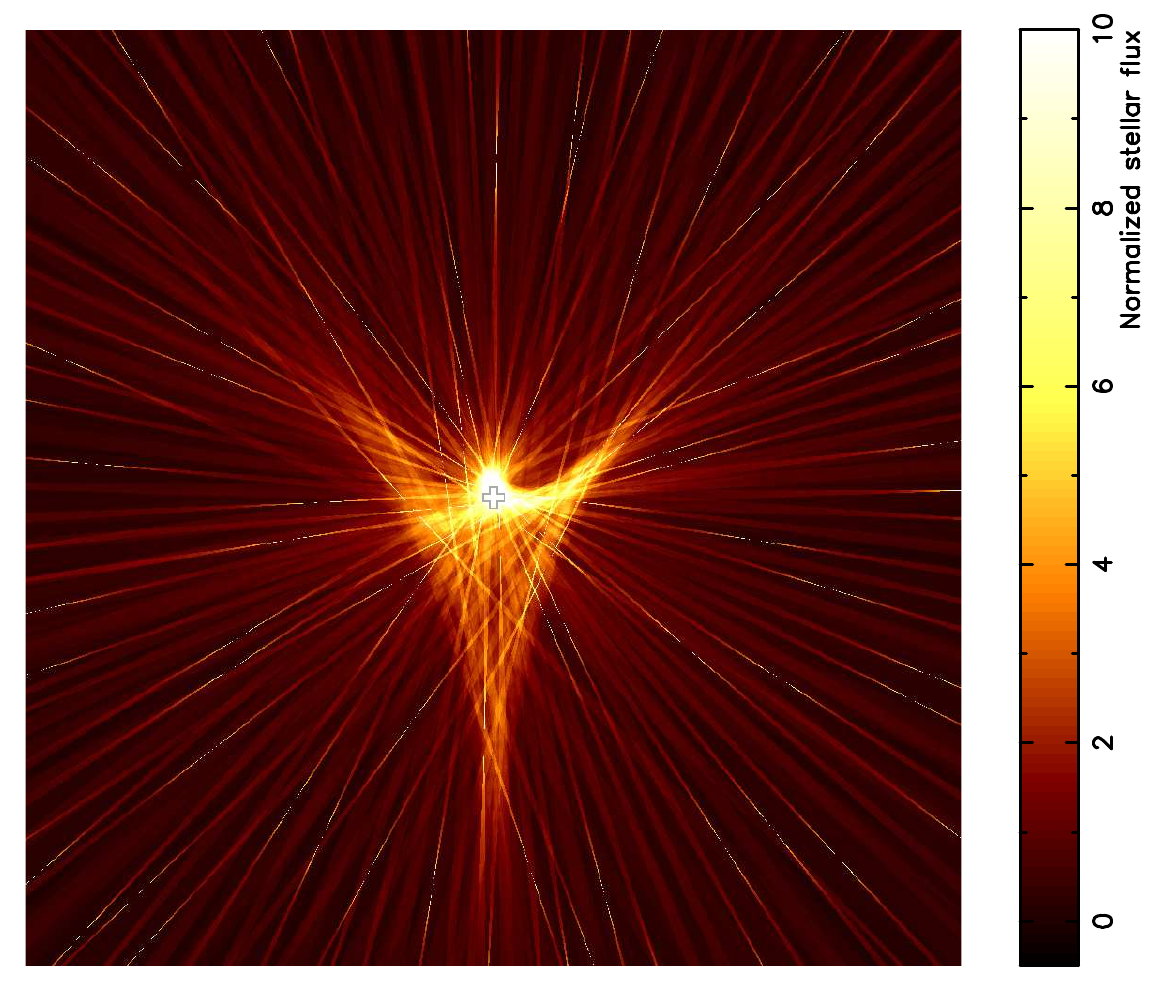}
\caption{%
\textit{Left:}
The same as the right panel of Fig.~\ref{fig_caustique_Titan_03}.
\textit{Right:} 
The effect of small vertical corrugations of the order of 100~m have been added to the general smooth
shape of the flash layer (left panel of  Fig.~\ref{fig_caustique_Titan_03}).
In the case of Titan, these corrugations are caused by fluctuations induced by atmospheric gravity waves. 
They cause many streaks in the flash region that result in flux fluctuations, or spikes, 
in the light curves (Fig.~\ref{fig_flash_Titan_03_Triton_17}).
}%
\label{fig_caustique_Titan_03_corrugated}
\end{figure}
%%%%%%%%%%%%%%%%%%%%%%%%%%%%%%%%%%%%%%%%%%

\subsection{Effect of atmospheric waves}

\begin{figure}[!t]
\centering
\includegraphics[totalheight=8cm]{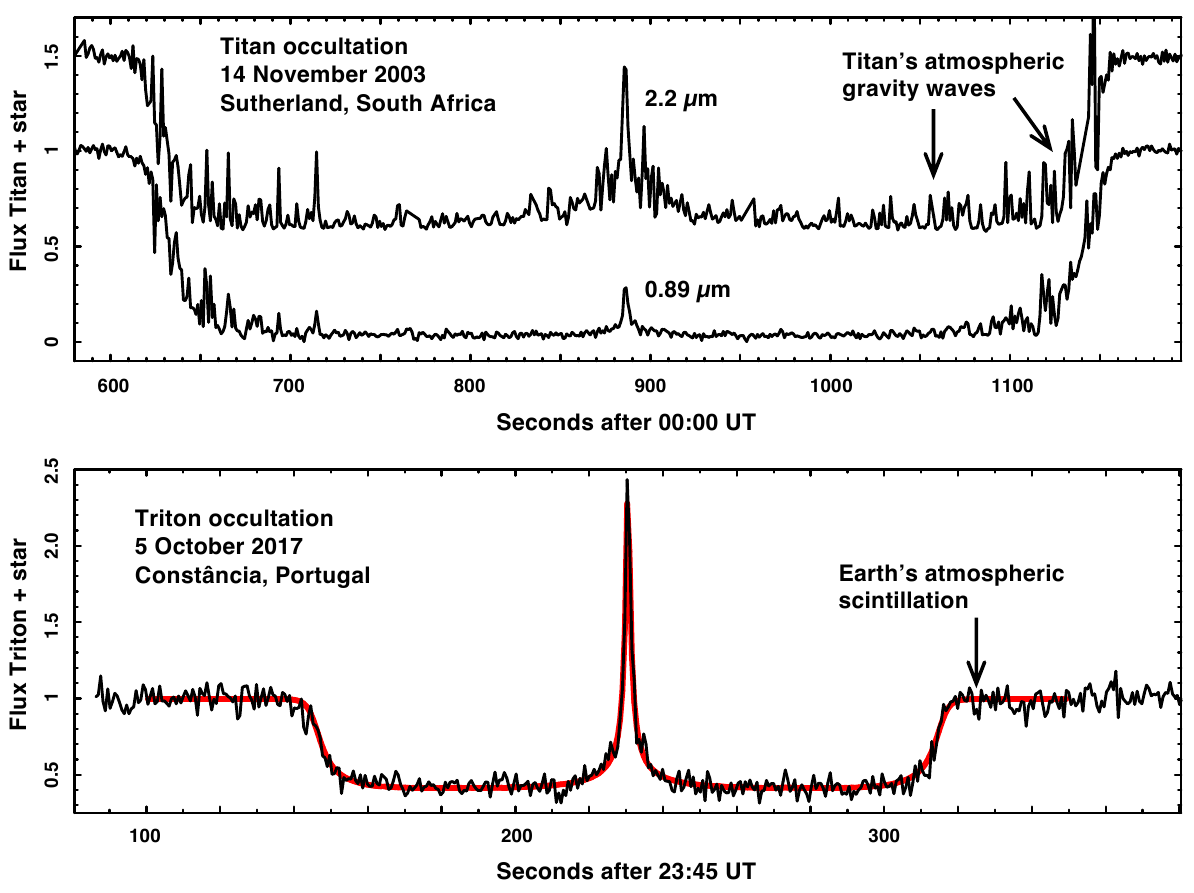}
\caption{%
\textit{Upper panel:}
the occultation by Titan's atmosphere observed on 14 November 2003 at the Sutherland in South Africa.
The lower  curve shows the event as observed in the visible (0.89~$\mu$m), 
while the upper curve shows the same event in the near infrared (2.2~$\mu$m, shifted vertically by +0.5 for a better viewing).
Conspicuous flux variations (or ``spikes") are seen everywhere during the occultation, including in the central flash.
The spikes are caused by gravity waves that propagate in Titan's atmosphere, 
creating in particular the streaks in the flash region (Fig.~\ref{fig_caustique_Titan_03_corrugated}). 
Note that the flash observed in the visible light is much weaker than its counterpart observed in the infrared.
This difference stems from absorption by hazes that are more opaque at 0.89~$\mu$m than at 2.2~$\mu$m \cite{sic06}.
\textit{Lower panel:} 
the occultation by Triton observed on 5 October 2017 at Const\^ancia in Portugal.
This station passed at a mere 6-km distance to Triton's shadow center.
The red line is a fit to the data assuming a spherical and transparent atmosphere.
The height of the flash represents more than three times the flux of the unocculted star,
a current record for this kind of observations.
The flux fluctuations seen in the light curve are caused by the Earth atmosphere, not by Triton's atmospheric waves
that are much weaker than for Titan.
This observation shows that, contrarily to Titan, Triton's atmosphere is essentially spherical and transparent \cite{mar22}. 
}%
\label{fig_flash_Titan_03_Triton_17}
\end{figure}

The left panel of Fig.~\ref{fig_caustique_Titan_03_corrugated} displays the intensity map of the Titan's flash
already shown in Fig.~\ref{fig_caustique_Titan_03}.
It stems from a flash layer which has a smooth profile.
Observations of various occultations by giant planets or Titan, however,  reveal irregular structures of the flash. 
More precisely, rapid fluctuations of the stellar flux (or ``spikes") are superimposed to the general smooth increase
of signal observe during the flash episode, see an example in Fig.~\ref{fig_flash_Titan_03_Triton_17}. 

These fluctuations are caused by internal gravity waves that propagate in Titan's upper atmosphere.
They create small ``corrugations" of the central flash layer that break down the smoothing varying centers of curvature 
of the limb into many centers of curvature.
This results into a blurring of the central flash intensity map.
In the case of Titan, these corrugations amount to some hundreds meters and cause the blurring
illustrated in Fig.~\ref{fig_caustique_Titan_03_corrugated}
\cite{sic06}.

%%%%%%%%%%%%%%%%%%%%%%%%%%%%%%%%%%%%%%%%%%
%%%%%%%%%%%%%%%%%%%%%%%%%%%%%%%%%%%%%%%%%%

\subsection{Opacity}

In Eq.~\ref{eq_phi_focusing}, we mentioned the existence of the factor $\exp(-\tau)$
which stems from the possible presence of an absorbing material in the atmosphere.
In fact, depending on the body, this term may become dominant compared to the effect of refraction.

An example of haze absorption is given in the upper panel of Fig.~\ref{fig_flash_Titan_03_Triton_17}.
A clear difference between the two flashes is observed, due to the differential extinction
between the I (visible) and the K (near infrared) 
bands\footnote{A difference also stems from the chromatic dependence of the refraction index,
but this effect is too small to be relevant here.}.
More precisely, the chromatic dependence of $\tau$ vs. wavelength is such that the atmosphere
is essentially transparent in the infrared, while being quite absorbant in the visible.

Since the flash strongly increases the flux when the star is deeply immersed in the atmosphere,
it is a useful tool to probe haze properties, and in particular its chromatic dependence.
As the zero stellar flux is usually ill-defined (due to the contribution of the occulting body), 
it is difficult to assess the haze optical depth outside the flash region,
where the residual stellar flux is small.

The other example of Fig.~\ref{fig_flash_Titan_03_Triton_17} is a flash observed during an occultation by Triton.
Contrarily to Titan, Triton's flash is completely explained (to within the noise in the data) by a spherical and transparent
atmosphere: in that sense, Triton's  atmosphere appears as a ``perfect lens". 

\section{Transits}
\label{sec_transits}

\subsection{Principle}

We now turn to the case where the occulted background star is angularly much larger than the foreground occulting body.
As mentioned in the Introduction, 
this  situation is described as a \textit{transit} (instead of an occultation). 
This occurs for instance when a exo-planet passes in front of its star,
or when the planet Venus is seen transiting in front of the solar disk.

\begin{figure}[!t]
\centering
\includegraphics[totalheight=5cm,trim=0 0 0 0]{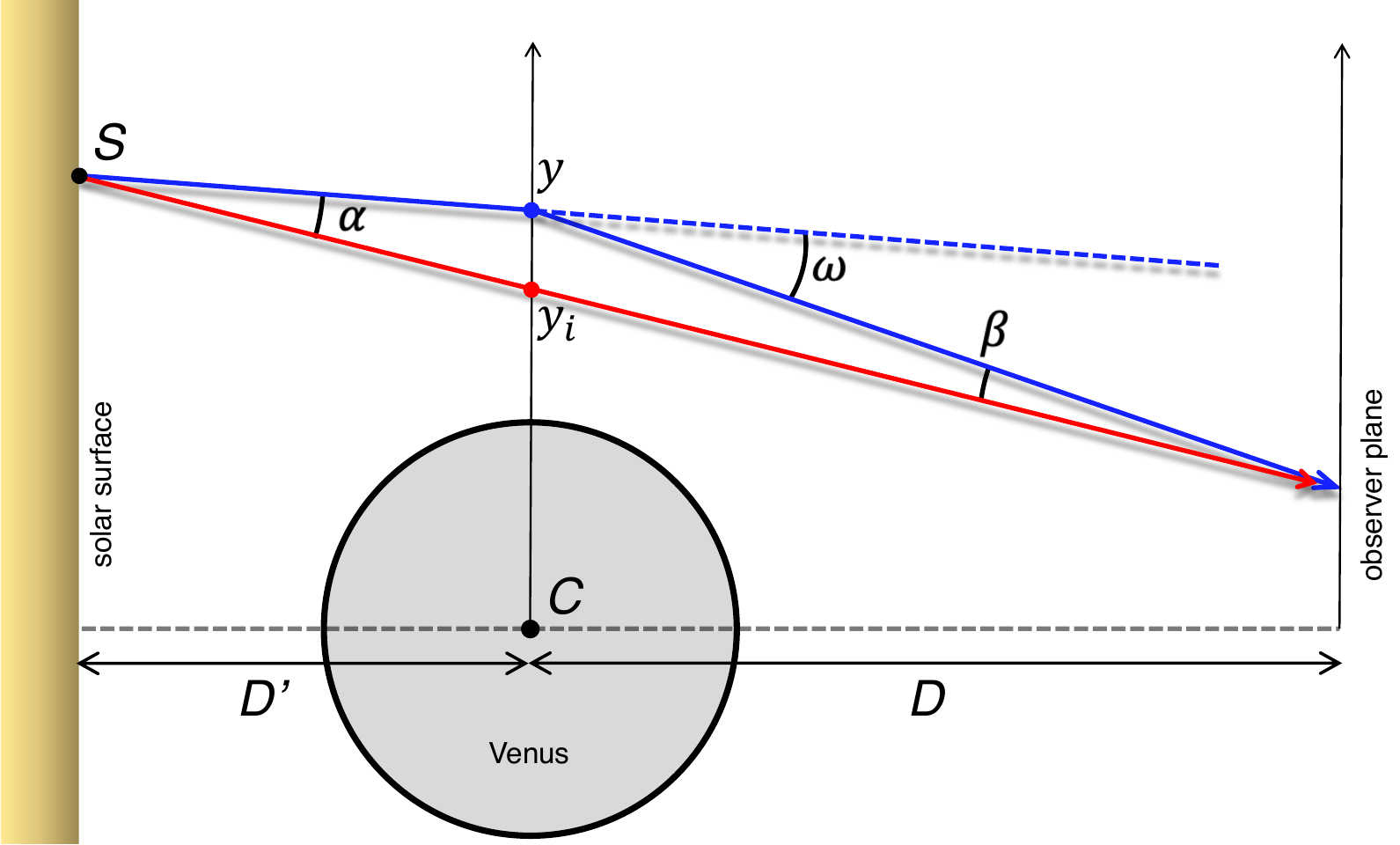}
\caption{%
The geometry of a transit phenomenon, with the definitions of the angles and other quantities used in the text.
The case examined here is the transit of Venus in front of the Sun. 
}%
\label{fig_lomo_geo_y}
\end{figure}

As an example, the geometry of a Venus transit is sketched in Fig.~\ref{fig_lomo_geo_y},
where $D$ and $D'$ are the distance of Venus to the Earth and to the Sun, respectively. 
Without refraction by Venus' atmosphere, an observer would receive a ray from a point $S$ on the Sun 
that intersects the plane perpendicular to the line of sight passing through Venus' center at ordinate $y_i$.
Due to refraction, however, the ray is deflected and appears to come from another ordinate $y$. 
In other words, the point $S$ that should be seen at $y_i$ has an image that is seen at $y$.

%%%%%%%%%%%%%%%%%%%%%%%%%%%%%%%%%%%%%%%%%%
%%%%%%%%%%%%%%%%%%%%%%%%%%%%%%%%%%%%%%%%%%

Taking by convention $\omega$ negative (as in Fig.~\ref{fig_deviation_total}) and
$\alpha$ and $\beta$ positive, we have $\omega= -(\alpha+\beta)$.
In the limit of small angles, we have $\alpha = (y-y_i)/D'$ and $\beta = (y-y_i)/D$, thus
$$
\omega = -\frac{D+D'}{DD'} (y-y_i).
$$

On the other hand, we can express $\omega$ as a function of $y$ (Eq.~\ref{eq_omega_vs_r}), so that
the equation above can be re-written
\begin{equation}
y = y_i + gH \exp \left[-\left(\frac{y-r_{1/2}}{H}\right)\right],
\label{eq_y_implicit_yi}
\end{equation}
which defines $r_{1/2}$ and
where $g$ is the dimensionless geometric 
factor\footnote{Not to be confounded with the acceleration of gravity $g(r)$ of the planet, see e.g. Eq.~\ref{eq_p_r}.}
$$
g= \frac{D'}{D+D'}.
$$

Eq.~\ref{eq_y_implicit_yi} provides implicitly $y$ as a function of $y_i$, 
that, is the position of the image of $S$ as seen in Venus' atmosphere by the observer. 
This equation can be solved numerically. 
Alternatively, we can introduce the quantity $\phi$ used before (but now as in auxiliary variable)
by defining it as $(1/\phi) - 1 = \exp [-(y-r_{1/2})/H]$ in analogy to 
(Eq.~\ref{eq_1_over_phi_ter}). From $y - y_i = y - r_{1/2} + r_{1/2} - y_i$, we  finally obtain
\begin{equation}
g \left( \frac{1}{\phi} - 1 \right) + \ln \left( \frac{1}{\phi} - 1 \right) = -\frac{y_i - r_{1/2}}{H}.
\label{eq_BC_transit_y}
\end{equation}
This is similar to the Baum and Code equation~\ref{eq_BC}, 
except from the appearance of the geometrical factor $g$ and 
from the fact that the term $(1/\phi - 2)$ has been replaced by $(1/\phi - 1)$.
We will refer to this equation as the ``modified Baum and Code equation".
For a given $y_i$, the inversion of the modified Baum and Code equation provides $\phi$,
which in turn yields $y$ through
\begin{equation}
y= y_i + gH \left( \frac{1}{\phi} -1 \right).
\label{eq_y_vs_yi}
\end{equation}

\begin{figure}[!h]
\centering
\includegraphics[totalheight=5cm,trim=0 0 0 0]{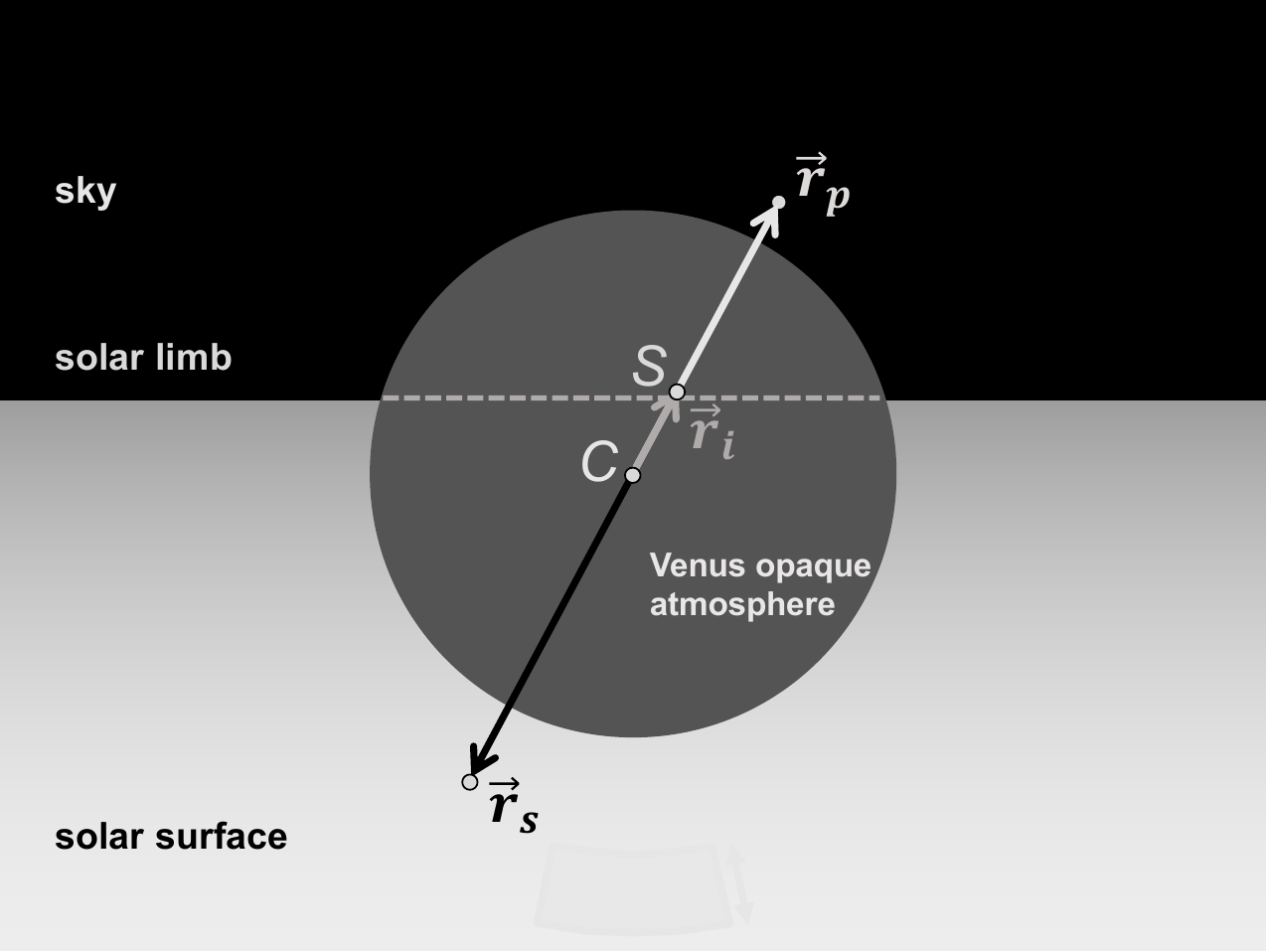}
\includegraphics[totalheight=5cm,trim=0 0 0 0]{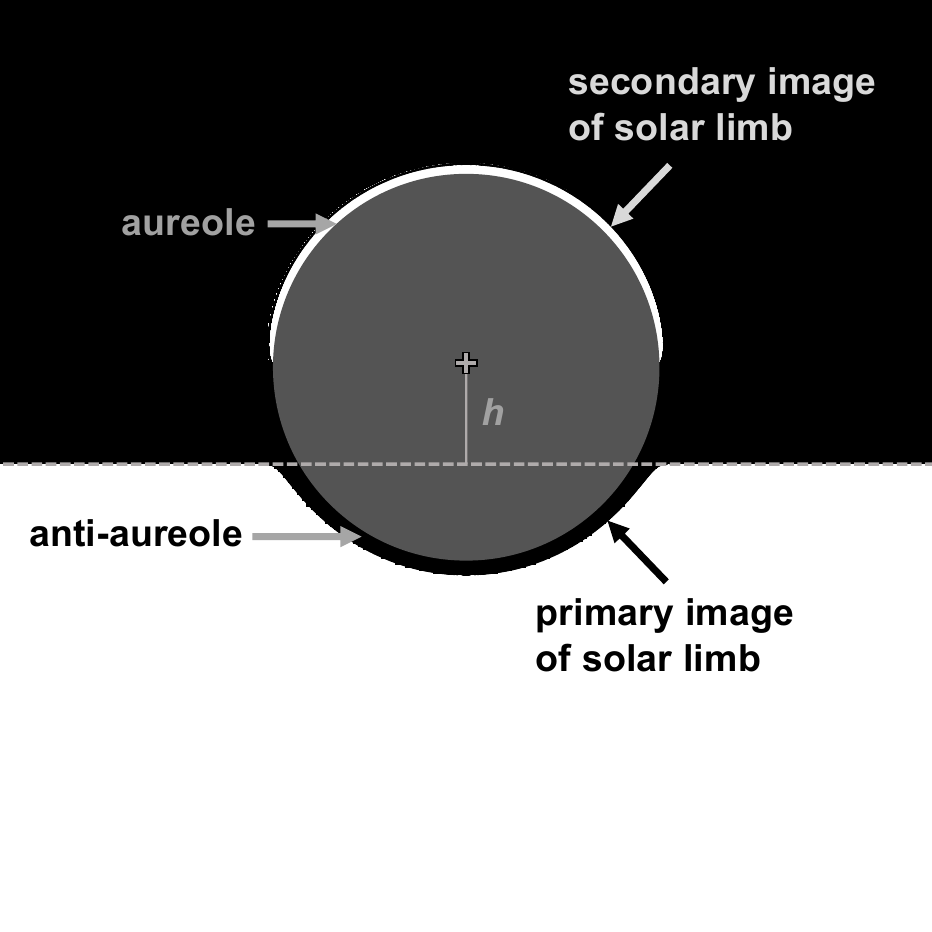}
\caption{%
\textit{Left:}
the point $S$ of the solar limb has two images after being refracted by the planet's atmosphere, 
assumed here to be spherical.
The primary (resp. secondary) image is located at $\vec{r}_p$ (resp. $\vec{r}_s$). 
\textit{Right:} 
the case shown here is the transit of Venus in front of the Sun, 
using here a radius of 6130~km for the opaque atmosphere \cite{tan12}. 
The quantity $h$ denotes the position of Venus' center (cross) relative the solar limb. 
By convention, $h$ is positive (resp. negative) if Venus' center is above (resp. below) the solar limb.
Two edgings appears along Venus' limb, an upper bright one (the ``aureole") and a lower dark one (the ``anti-aureole").
They are obtained by using Eqs.~\ref{eq_r_lomo} to
calculate the primary and secondary images of the solar limb.
If Venus' center were below the projected solar limb (as it is in the left panel),
then the primary and secondary nature of the images would be swapped.
Assuming a transparent atmosphere, the conservation of radiance implies that 
the aureole has the same brightness as the Sun.
Likewise, the anti-aureole has the same brightness as the background sky, in black color here.
For a better viewing of the aureole and the anti-aureole, 
the scale height $H$ of Venus' atmosphere used here has been largely exagerated (100~km) 
compared to the actual value of about 4~km \cite{tan12}. 
Also, Venus' disk is plotted here in gray for better visibility, but it is by no way luminous in actual observations.
It has actually the same brightness as the background sky.  
}%
\label{fig_lomo_geo_r}
\end{figure}

In this equation, we can choose without loss of generality $y>0$. In this case,
$y_i > 0$ (resp. $y_i < 0$) corresponds to the primary (resp. secondary) image of $S$. 
More generally, this equation can be used to relate the vector position $\vec{r}_i$ of $S$
projected at Venus to the vector position $\vec{r}$ of its image (Fig.~\ref{fig_lomo_geo_r}).
Denoting $r_i= || \vec{r}_i ||$, we can re-write Eq.~\ref{eq_y_vs_yi} in a vectorial form.
We can encapsulate in the same equations the cases of the primary images and secondary images.
This is done by defining a parameter $\epsilon=+1$ (resp. $\epsilon=-1$) for primary (resp. secondary) images. 
Then we pose 
\begin{equation}
\Delta r = \epsilon r_i - r_{1/2} {\rm \ \ and \ \ } u = \frac{1}{\phi} - 1,
\label{eq_Delta_r}
\end{equation}
so that the equations relating $\vec{r}$ and $\vec{r}_i$ are
\begin{equation}
\left\{
\begin{array}{l}
\displaystyle
g u + \ln(u) = -\frac{\Delta r}{H} \\
 \\
\displaystyle
\vec{r}= \left( 1 + \frac{gHu}{\epsilon r_i} \right) \vec{r}_i, \\
\end{array}
\right.
\label{eq_r_lomo}
\end{equation}

In numerical schemes, one can get rid of the factor $\epsilon$ in Eqs.~\ref{eq_Delta_r} and \ref{eq_r_lomo}
by adopting $r_i > 0$ for primary images and $r_i < 0$ for secondary images.
This trick must be used with care, however, as $r_i$ classically denotes the modulus of $\vec{r}_i$ 
and thus, is in principle always positive.

For a given $r_i$, the first equation of the system \ref{eq_r_lomo} (i.e. the  modified Baum and Code equation) provides $u$. 
Once this is done, the second equation provides the vector position $\vec{r}$ of the image of the point $S$ located at $\vec{r}_i$.
For $D' \rightarrow +\infty$, we have $g=1$, 
and the second equation of the system above reduces to Eq.~\ref{eq_r_vs_z}, as expected.
 
It is instructive to consider the asymptotic behavior of $\vec{r}$ for $\Delta r/H$ approaching $-\infty$,  
corresponding to $\phi$ approaching zero (or equivalently, $u$ approaching $+\infty$), 
i.e. images of $S$ that are deeply immersed into the atmosphere,
referred to as the ``deep image regime" hereafter.

The modified Baum and Code equation provides a first estimation $u \sim - \Delta r/gH$ 
by neglecting $\ln(u)$ with respect to $u$.
Introducing this expression of $u$ back into the modified Baum and Code equation, 
we obtain $gHu \sim -\Delta r - H \ln(-\Delta r/gH)$.
Finally, using this approximation of $gHu$ in the second equation of the system~\ref{eq_r_lomo}, we see that 
in the deep image regime, the image is located at distance
\begin{equation}
r \sim  r_{1/2} - H \ln \left( \frac{-\Delta r}{gH} \right)
\label{eq_r_deep}
\end{equation}
from the planet center.

Because of the weak logarithmic dependence, even for large negative values of $\Delta r$, 
the deep image of $S$ probes atmospheric layers  that are only a few scale heights below the half light radius $r_{1/2}$,
At this point, other effects can dominate the aspect of the image.
For instance, if the atmosphere is tenuous enough, the image may hit at some point the surface of the planet and merely disappears.
In some other cases, the deep atmosphere may be hazy or cloudy, and the image may enter opaque regions,
causing also its disappearance.
This point is now discussed further in the particular case of Venus transits. 

%%%%%%%%%%%%%%%%%%%%%%%%%%%%%%%%%%%%%%%%%%
%%%%%%%%%%%%%%%%%%%%%%%%%%%%%%%%%%%%%%%%%%

\subsection{The Lomonossov effect}

The discovery of Venus' atmosphere is traditionally  attributed to Mikhail Lomonossov, who observed 
the Venus transit of 6 June 1761 from St Petersburg Observatory \cite{mar05}.
The credit of this original discovery by Lomonossov is debated, though \cite{lin69,pas11,pas12,tan12}.
However, Lomonossov's basic interpretation was correct, that is, the luminous ring appearing along Venus' limb
as the planet emerges from the apparent solar disk is caused by atmospheric refraction.

This effect should not be confused with the extension of Venus' crescent near inferior conjunction,
caused by the quasi forward-scattering of light by hazes, another evidence that Venus does possess an atmosphere.
This point is discussed in the next subsection, and
we will see that forward-scattering plays a negligible role during Venus transits, when compared to refraction.

\begin{figure}
\centering
\includegraphics[totalheight=7cm,trim=0 0 0 0]{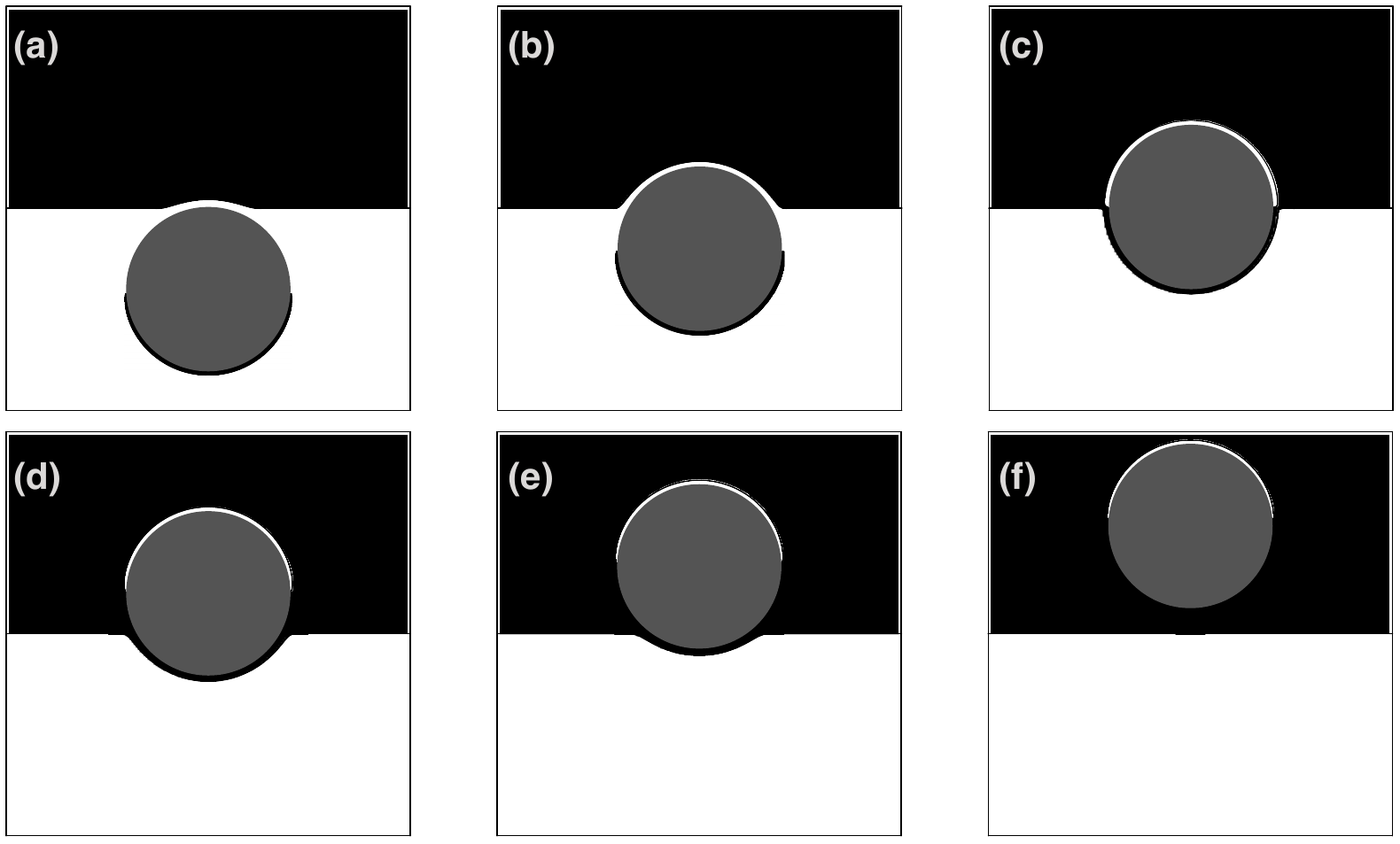}
\caption{%
A sequence of images showing the evolution of the aureole and anti-aureole 
as Venus emerges from the apparent solar disk.
In the various images, the projected Venus' center is respectively located at
6000 and 3000~km below the solar limb (panels (a) and (b)), and at
100, 3000, 5000 and 8000~km above the solar limb (panels (c), (d), (e) and (f)).
Note that the aureole and the anti-aureole have the same width when Venus' center is aligned with the solar limb (panel (c)).
When Venus is far down in front of the Sun, the anti-aureole disappears behind the opaque Venus' atmosphere.
The same occurs with the aureole when Venus is far up in front of the background sky. 
}%
\label{fig_lomo_m6000_to_p8000_km}
\end{figure}

The Lomonossov effect is visualized by calculating the primary and secondary images of the point $S$
as this point is moved along the solar limb (Fig.~\ref{fig_lomo_geo_r}).
The time series of Fig.~\ref{fig_lomo_m6000_to_p8000_km} shows in more details the evolution of the two ``edgings"
resulting from the Lomonossov effect.
The bright edging (called the ``aureole" for short hereafter) is always outside the solar limb, 
while the dark edging (called the ``anti-aureole") is always inside the solar limb. 
In practice, the anti-aureole goes unnoticed in the case of Venus, as it is confounded with the planet dark side.
However, it must be accounted for when simulating transit light curves involving  exo-planets.
For instance, the panel (c) of Fig.~\ref{fig_lomo_m6000_to_p8000_km} shows that when the planet center
projects itself on the solar limb, the contributions of the aureole and the anti-aureole cancel out.
Then, the flux taken away by the planet just corresponds to the flux blocked by its own disk. 

%%%%%%%%%%%%%%%%%%%%%%%%%%%%%%%%%%%%%%%%%%
%%%%%%%%%%%%%%%%%%%%%%%%%%%%%%%%%%%%%%%%%%

The visibilities of the aureole and anti-aureole depend on the value of $r_{\rm opa}$, the radius of the opaque atmosphere.
The image of a given point $S$ on the solar limb is located at a distance $r$ of Venus' center which is given by
Eq.~\ref{eq_r_deep}, i.e. $r  \sim  r_{1/2} - H \ln ( -\Delta r/gH )$.
This image is effectively observed if $r > r_{\rm opa}$. 
Using the definition of $\Delta r$ (Eq.~\ref{eq_Delta_r}), the condition of visibility is
\begin{equation}
\epsilon r_i > r_{1/2} - gH \exp \left( \frac{r_{1/2} - r_{\rm opa}} {H} \right).
\label{eq_bright_edging_visibility}
\end{equation}

\begin{figure}
\centering
\includegraphics[totalheight=5cm,trim=0 0 0 0]{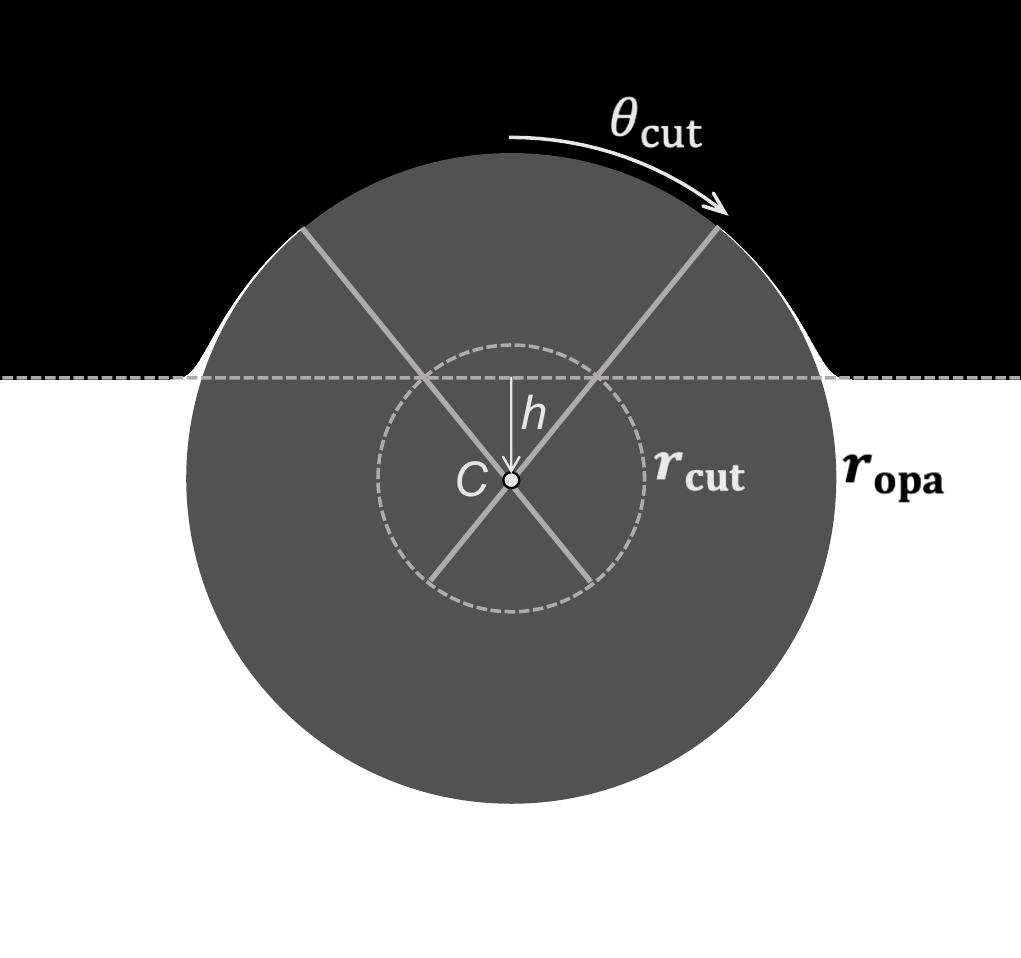}
\includegraphics[totalheight=5cm,trim=0 0 0 0]{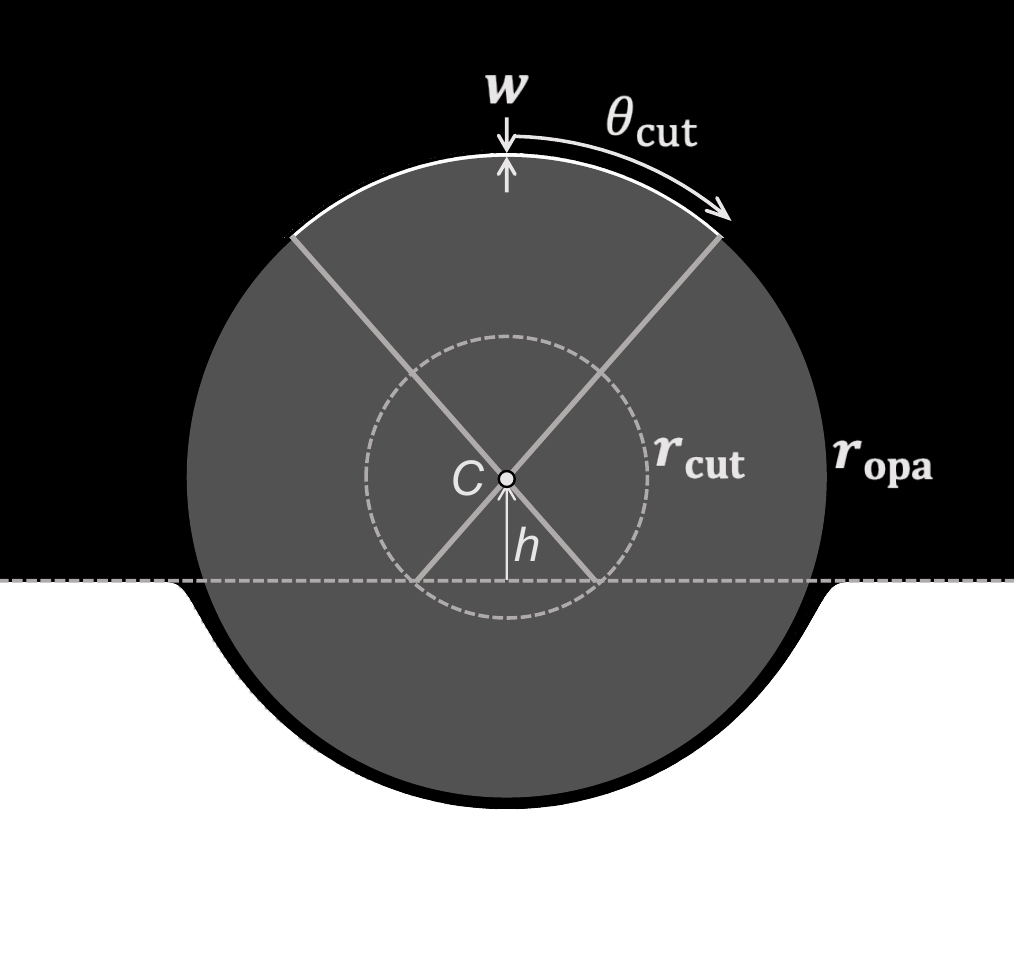}
\caption{%
\textit{Left:}
condition of visibility for the aureole for $h <0$, i.e. when Venus' center is below the solar limb (Eq.~\ref{eq_bright_edging_visibility_prima}).
The aureole disappears behind the opaque atmosphere if its angle with the vertical is less than $\theta_{\rm cut}$.
\textit{Right:}
the reversed situation for  $h > 0$. Now the aureole is visible when its angle with the vertical is less than $\theta_{\rm cut}$.
In both cases, $\theta_{\rm cut}= \arccos ( |h|/r_{\rm cut} )$, see text for details.
}%
\label{fig_lomo_liseres_r_cut}
\end{figure}

We can consider two cases, depending on the sign of $h$, the height of Venus' center above the solar limb, see Fig.~\ref{fig_lomo_liseres_r_cut}.
For $h < 0$ (left panel of Fig.~\ref{fig_lomo_liseres_r_cut}), the aureole is the primary image of the limb, so that $\epsilon= +1$,
and the condition of visibility of the aureole is
\begin{equation}
r_i > r_{\rm cut} = r_{1/2} - gH \exp \left( \frac{r_{1/2} - r_{\rm opa}} {H} \right),
\label{eq_bright_edging_visibility_prima}
\end{equation}
where we define the cutoff value $r_{\rm cut}$.
 
Geometrical considerations based on the examination of Fig.~\ref{fig_lomo_liseres_r_cut} (left panel)
show that the aureole  is visible if its angle with the vertical is larger than the cutoff angle
$$
\theta_{\rm cut}= \arccos \left( \frac{|h|}{r_{\rm cut}} \right),
$$
As $r_{\rm opa}$ decreases, $r_{\rm cut}$ decreases as well, until it reaches the value of $|h|$.
At this point, 
and for all values of $r_{\rm cut}$ between 0 and $h$, 
we have $\theta_{\rm cut} = 0$ and the aureole is uninterrupted along the upper limb of Venus,
as illustrated for instance in panel (b) of Fig.~\ref{fig_lomo_m6000_to_p8000_km}.

For $r_{\rm cut} = 0$, the aureole is complete when $h=0$ and extends over $\pi$ radians, 
a situation illustrated in panel (c) of Fig.~\ref{fig_lomo_m6000_to_p8000_km}.
A new regime sets in for $h > 0$, as the aureole now corresponds to secondary images of the solar limb ($\epsilon = -1$),
so that Eq.~\ref{eq_bright_edging_visibility} reads
\begin{equation}
r_i < r_{\rm cut} = -r_{1/2} + gH \exp \left( \frac{r_{1/2} - r_{\rm opa}} {H} \right).
\label{eq_bright_edging_visibility_secon}
\end{equation}
This situation is depicted in the right panel of Fig.~\ref{fig_lomo_liseres_r_cut}.
The aureole now extends over an angle $2 \theta_{\rm cut}$, where again we have $\theta_{\rm cut}= \arccos ( |h|/r_{\rm cut} )$.

%%%%%%%%%%%%%%%%%%%%%%%%%%%%%%%%%%%%%%%%%%
%%%%%%%%%%%%%%%%%%%%%%%%%%%%%%%%%%%%%%%%%%

As $r_{\rm opa}$ decreases, $r_{\rm cut}$ increases until it is larger than the planet diameter.
Then, the aureole remains visible even if the planet disk is completely detached from the solar disk.
This is indeed the case for Venus. 
During the 8 June 2004 transit, 
estimates of the various quantities entering the expression of $r_{\rm cut}$ were derived \cite{tan12}:
$H \sim 4$~km,  $r_{1/2} \sim 6170$~km and $r_{\rm opa} \sim 6130$~km.
Using the geometrical factor $g=0.716$ relevant to that event, 
we obtain $r_{\rm cut} \sim 57,000$~km.
This means that the aureole could be observed 
even in Venus' disk is completely detached from the solar disk, 
a situation illustrated in panel (f) of Fig.~\ref{fig_lomo_m6000_to_p8000_km}.

Beyond the distance of 57,000~km, the aureole disappears behind Venus' opaque atmosphere.
As Venus is at $\sim 0.3$~au from Earth under these circumstances, 
this corresponds to an angular separation of a mere 4 arcmin between Venus and the solar limb.

The equation $\theta_{\rm cut}= \arccos ( |h|/r_{\rm cut} )$ 
shows that for $r_{\rm cut}$ very large, $\theta_{\rm cut}$ approaches $\pi/2$, but it cannot goes beyond this value. 
In other words, the aureole can be observed only along the Venus limb opposite to the Sun, and cannot exceed an angular extension of $\pi$ radians.
This is expected from the fact that solar rays passing near the lower limb of Venus  
(panel (f) of Fig.~\ref{fig_lomo_m6000_to_p8000_km})
cannot be refracted back to Earth, as the curvature of the ray is always pointing to the direction of increasing refractive index (Eq.~\ref{eq_dtau_ds}).

Another aspect, however, limits the observation of the aureole. Even though the brightness of the luminous arc is very
high (it is actually the brightness of the solar surface), it becomes very narrow for large values of $r_i$.
The maximum width $w$ of the aureole is reached at the top of Venus' limb (Fig.~\ref{fig_lomo_liseres_r_cut}), where $r_i = h$.
Thus, the maximum width reached by the aureole is $w= r - r_{\rm opa}$, 
where $r$ is given by Eq.~\ref{eq_r_deep}, with $\Delta r= -h - r_{1/2}$, so that
\begin{equation}
w \sim  r_{1/2} - r_{\rm opa} - H \ln \left( \frac{h+ r_{1/2}}{gH} \right),
\label{eq_width_edging}
\end{equation}
where we recall that here $h>0$.
As an example, we take $h= r_{\rm opa}$, corresponding to the situation where Venus' disk is tangent to the solar limb
(called the first and fourth contacts in the terminology of Venus' transits).
Adopting the numerical values mentioned before, we obtain $w \sim 6.5$~km.
This width cannot be resolved using any imaging technique from Earth.
Under usual seeing conditions, the Point Spread Function (PSF) due to our atmosphere of typically one arcsec.
This corresponds to about 210~km projected at the planet, which was at 0.288~au from Earth on 8 June 2004.
In the example taken above, the PSF dilutes the apparent brightness of the aureole by a factor of $210/6.5 \sim 30$.
As Venus' disk recesses away from the solar disk, the width $w$ of te aureole decreases and the dilution factor increases. 
At some point, the aureole, although intrisincally very bright, becomes too narrow, and thus too faint, to be detected.

\begin{figure}
\centering
\includegraphics[totalheight=5cm,trim=0 0 0 0]{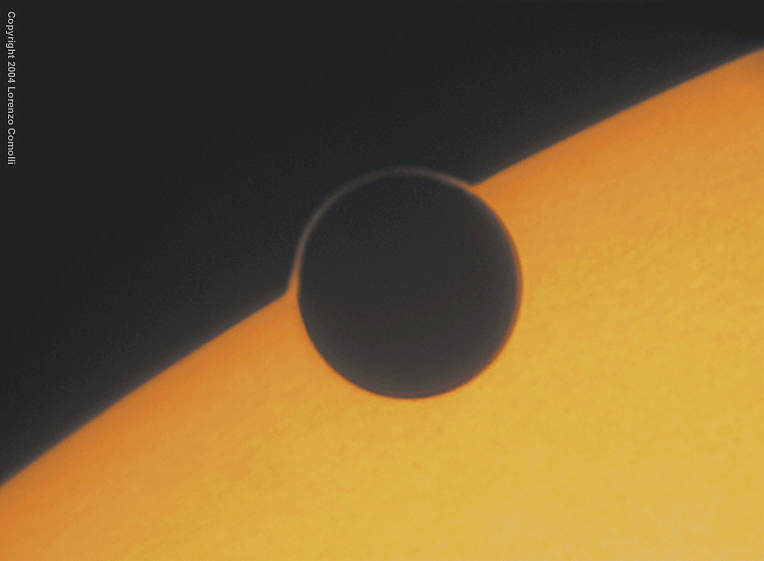}
\includegraphics[totalheight=5cm,trim=0 0 0 0]{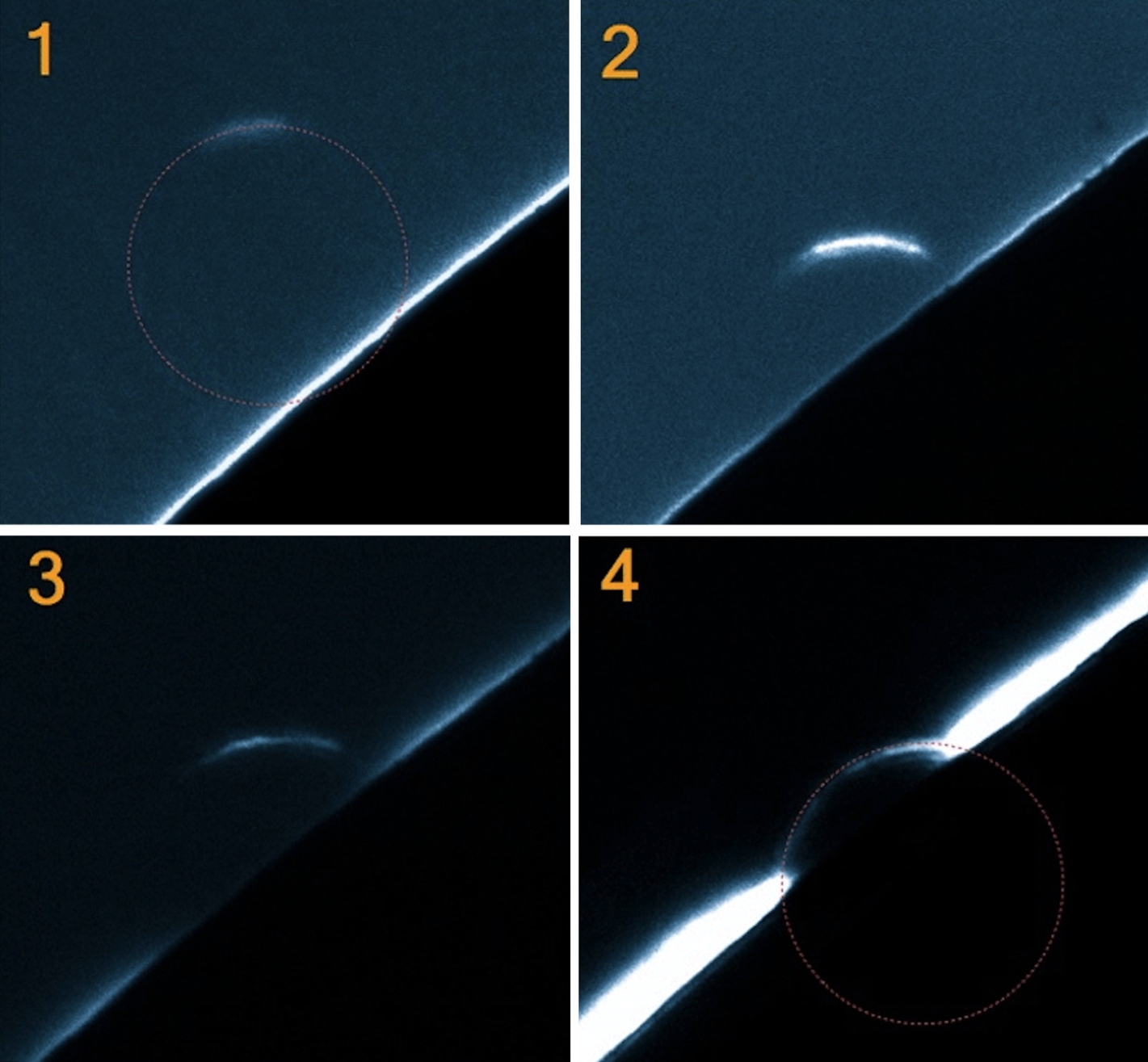}
\caption{%
\textit{Left:}
The Lomonossov aureole observed from Tradate (Italy) during the Venus transit of 8 June 2004,
see www.astrosurf.com/comolli/cong5e.htm (copyright Lorenzo Comolli).
It can be compared to the synthetic time series displayed in Fig.~\ref{fig_lomo_m6000_to_p8000_km}.
\textit{Right:}
The passage of Venus in front on the solar limb during the transit of 5/6 June 2012,
observed from Lowell Observatory in Arizona, USA (copyright Paolo Tanga).
One can note the irregular aspect of the aureole, possibly due to the varying altitude 
of the cloud deck in Venus' atmosphere. See text for discussion.
}%
\label{fig_Venus_transit_images}
\end{figure}

In the particular case of Venus' transits observed from Earth, 
the particular values of $r_{1/2}$, $r_{\rm opa}$ and $H$ are such that
 $w$ in Eq.~\ref{eq_width_edging} is (coincidentally) quite small. 
Thus, a small variation of $r_{\rm opa}$ induces a large relative variation of $w$.
This creates a patchy aspect of the aureole (Fig.~\ref{fig_Venus_transit_images})
and is a way to map the altitude of the cloud deck along Venus' limb.

%%%%%%%%%%%%%%%%%%%%%%%%%%%%%%%%%%%%%%%%%%
%%%%%%%%%%%%%%%%%%%%%%%%%%%%%%%%%%%%%%%%%%

The same kind of behaviors occur symmetrically for the anti-aureole.
In particular, if Venus' disk is deep inside the solar disk, it disappears behind the opaque layer.
Then, it is very difficult by direct imaging to know that the planet has an atmosphere, 
since the brightness of the Sun, even when observed through the atmosphere, remains unchanged.
In this case, the only way to detect the atmosphere is to observe the passage of Venus in front of a sunspot.
The image of the spot will be distorted in the same as a stellar disk is distorted when observed through the atmosphere (Fig.~\ref{fig_prima_secon_grazing}).
This would be an alternative to study Venus' atmosphere. However, such an event has a low probability to occur.
Note that another way to detect Venus' atmosphere would be to perform spectroscopic transit observations to detect 
the gaseous CO$_2$.

As a final remark, 
we note that the general formalism developed here can be applied to exoplanets transiting in front of their stars.
Some detailed calculations and applications to exoplanetary atmospheres are given in \cite{hub01}.

\subsection{Refraction vs. forward-scattering}

\begin{figure}
\centering
\includegraphics[totalheight=4cm,trim=0 0 0 0]{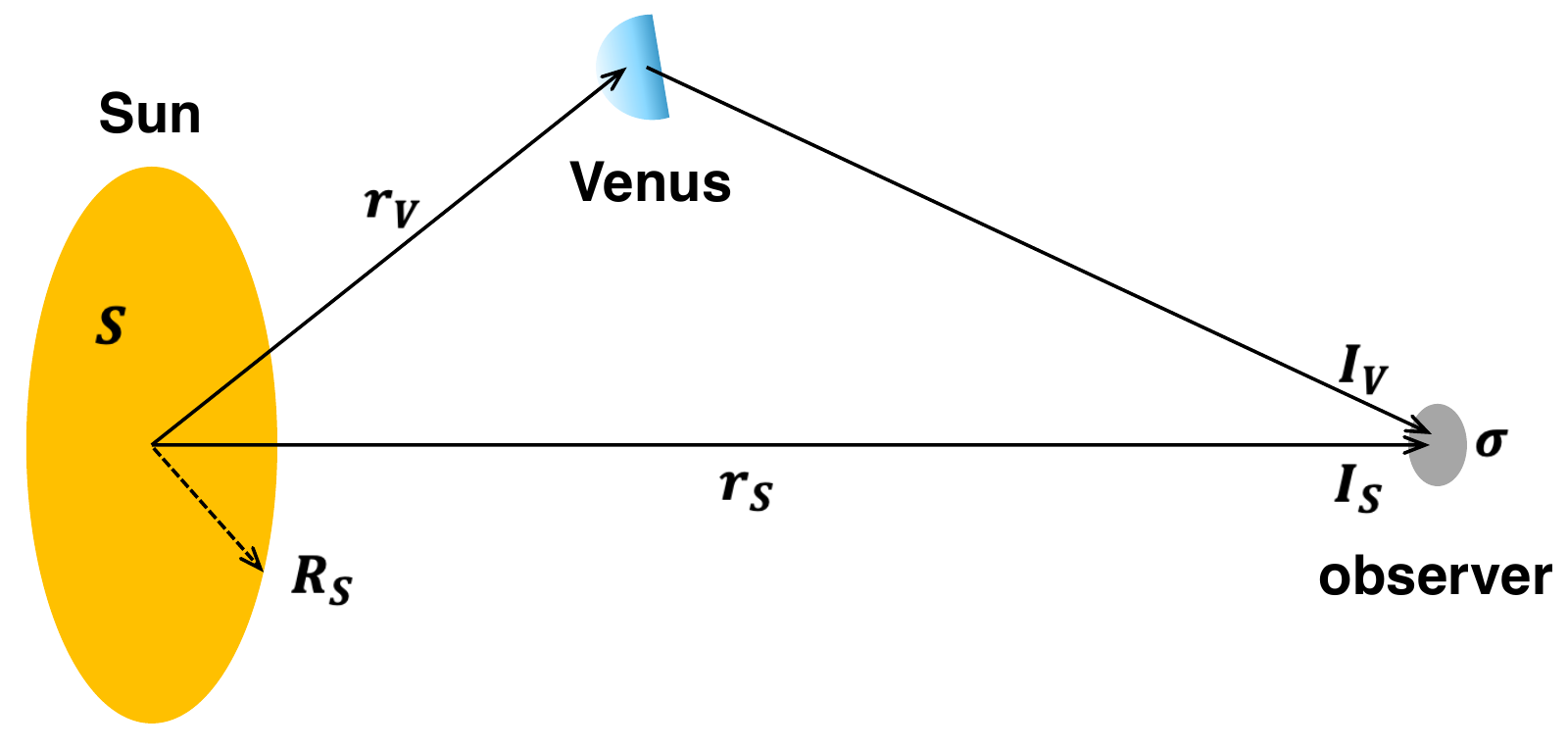}
\includegraphics[totalheight=4cm,trim=0 0 0 0]{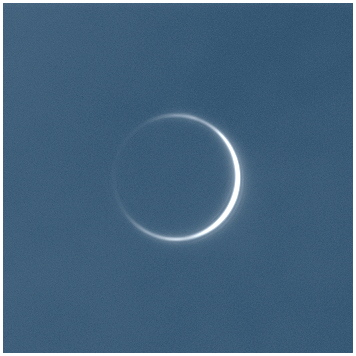}
\caption{%
\textit{Left:}
The geometry of light scattered by Venus back to Earth. See text for details.
\textit{Right:}
The extension of Venus' crescent caused by haze forward-scattering.
This picture was taken on 2 June 2020, as Venus was at about two degrees of the Sun within one degree of the Sun during inferior conjunction (Copyright Thierry Legault).
% \cite{law20},
% see also https://apod.nasa.gov/apod/ap200608.html.
% Copyright Pete Lawrence? Digitalsky http://digitalsky.org.uk
}%
\label{fig_I_forward_scatter}
\end{figure}

Near inferior conjunction, Venus' crescent extends beyond 180 degrees,
the value expected for an opaque, airless planet. 
This was first reported by Johann Schr\"oter in 1790, three decades after Lomonossov's observations \cite{sch96},
and then by various observers during the nineteenth century, see the review by Henry Norris Russell in 1899 
\cite{rus99}\footnote{This was the first paper published by Russell, at the age of twenty-two, 
well before he became famous for his work on stellar classification, which led to the Hertzsprung-Russell diagram.}
and the example displayed in Fig.~\ref{fig_I_forward_scatter}. 

Russell considered refraction and haze scattering as a possible causes for the extension of Venus' crescent.
He then inferred (rightly) that such extension was likely caused by haze scattering rather than refraction.
This is in line with the results obtained in the previous subsection: as soon as Venus projects itself at more than a few 
arcmin from the solar limb, the ``refractive aureole" disappears due to the presence of an opaque layers in the atmosphere,
leaving only a ``haze aureole".
 
However, during a transit the aureole is largely dominated by refraction, not by haze scattering.
Let us consider for this the solar disk with radius $R_S$ and apparent surface area $S= \pi R_S^2$, 
emitting rays with radiances (or intensities) $I_S$ towards an observer at distance $r_S$, 
who receives the solar flux over a surface area $\sigma$ (Fig.~\ref{fig_I_forward_scatter}).
Ignoring factors of order unity that account for the angles of emission or reception, 
the luminous power received from the Sun at $\sigma$ is $\Phi_S= I_S S \sigma/r_s^2$, 
so the solar flux $f_S$ (also called the solar constant) is $f_S= \Phi_S/\sigma=I_S S/r_s^2$.
Let us denote $f_V$ the solar constant at Venus,
at distance $r_V$ from the Sun.
Then, Venus' lit hemisphere re-emits towards $\sigma$  rays with radiance $I_V= p \Phi(\alpha) f_V/\pi$.
Here, $p$ is the geometric albedo and $\Phi(\alpha)$ is the phase function of the atmosphere, 
where $\alpha$ is the angle Sun-Venus-observer.
Consequently, the ratio $I_V/I_S$ is
\begin{equation}
\frac{I_V}{I_S} =  p \Phi(\alpha) \left( \frac{R_S}{r_V} \right)^2.
\label{eq_IV_over_IS}
\end{equation}

As $R_S \sim 700,000$~km and $r_V \sim 0.7$~au $\sim 10^8$~km, 
and because $p \Phi(\alpha)$ is of order unity,
we obtain $I_V/I_S < 10^{-2} \ll 1$.
As mentioned before, $I_S$ is also the brightness of the refractive aureole, 
while $I_V$ is the brightness of the lit side of Venus, as seen from the surface element $\sigma$.
The radiance (or brightness) on the rays forward-scattered by hazes to Earth is itself much less than $I_V$,
so that the brightness of the haze aureole is several orders of magnitude fainter than
the brightness of the refractive aureole.
This makes impossible the detection of the haze aureole, for instance along the limb that is nearest to the Sun when Venus' disk is
completely outside the solar disk (panel (f) of Fig.~\ref{fig_lomo_m6000_to_p8000_km}).

%%%%%%%%%%%%%%%%%%%%%%%%%%%%%%%%%%%%%%%%%%
%%%%%%%%%%%%%%%%%%%%%%%%%%%%%%%%%%%%%%%%%%

More discussion on the respective roles of refraction and scattering in the case of exoplanet transits 
is provided in \cite{hub01}.
In particular, we note that in exoplanetary cases, 
the solar radius $R_S$ and Venus heliocentric distance $r_V$ in Eq.~\ref{eq_IV_over_IS}
are replaced by the stellar radius $R_\star$ and the planet distance $r_P$, respectively.
As $R_\star$ and $r_P$ may be comparable, the contribution of forward-scattering relative to
refraction may be not overwhelmingly small, and thus should be considered when generating
synthetic transit light curves. 

\section{Conclusion}

The various aspects of refraction during stellar occultations and transits described in this chapter
illustrate the long-standing interest of the astronomical community in refraction phenomena.
Transits permitted to discover Venus' atmosphere in the eighteenth century, 
and occultations revealed Pluto's atmosphere in the years 1980's, thus paving the way to
the NASA New Horizons flyby of the dwarf planet in 2015. 

Meanwhile, occultations and transits still continue to raise great interest. 
They are used to monitor long-term seasonal evolutions of the atmospheres of Titan, Triton and Pluto, among others. 
They also probe subtle dynamical effects such as gravity waves that are just impossible to track using any other
Earth-based methods.

Occultations and transits have a bright future in store for us. 
The discovery of tenuous atmospheres around remote Trans-Neptunian  Objects
will only be possible by observing stellar occultations, to a sensitivity as good as a few nanobars.
Also, time series such the one displayed in Fig.~\ref{fig_lomo_m6000_to_p8000_km} can be applied
to the study of atmospheres of exoplanets. 
By comparing transit photometric light-curves with models, 
one can constrain key parameters such as the scale height, cloud altitude and density profiles 
of the atmospheres of these remote worlds.   

\vspace{5mm}
\noindent
\textbf{Acknowledgements} \\

\noindent
I thank Luc Dettwiller and an anonymous reviewer for corrections and constructive comments.

%%%%%%%%%%%%%%% BIBLOGRAPHY %%%%%%%%%%%%%%%%%%%%%%%%%
% The next command determines the bibliography style. Please do not
% change this.
\bibliographystyle{crunsrt}

%This calls all references from the .bib
%\nocite{*}

%  This inserts the bib file
\bibliography{crphys_Bruno}
%%%%%%%%%%%%%%%%%%%%%%%%%%%%%%%%%%%%%%%%%%%%%%%%%

\end{document}